\documentclass[final,3p,times]{elsarticle}

\usepackage[OT1]{fontenc}
\usepackage{amsmath,mathtools}
\usepackage{graphicx}
\usepackage{amssymb}
\usepackage{upquote}
\usepackage{ytableau}
\ytableausetup{smalltableaux}
\usepackage{breqn}
\usepackage{cadabra}
\usepackage{dirtree}
\usepackage{relsize}
\biboptions{compress}

\def\ContinueLineNumber{\lstset{firstnumber=last}}
\def\StartLineAt#1{\lstset{firstnumber=#1}}

\newcommand{\Lag}{\mathcal{L}}

\journal{Computer Physics Communications}
\usepackage[colorlinks]{hyperref}

\begin{document}

\begin{frontmatter}
  \title{Cadabra and Python algorithms in General Relativity and
    Cosmology I: Generalities}

  \author[1]{Oscar Castillo-Felisola}
  \ead{o.castillo.felisola@protonmail.com}


  \author[2]{Dominic T. Price\corref{cor1}}
  \ead{dominicprice@outlook.com}

  \author[3]{Mattia Scomparin}
  \ead{mattia.scompa@gmail.com}
  \cortext[cor1]{Corresponding author}

  \address[1]{Departamento de F\'isica and Centro Cient\'ifico y
    Tecnol\'ogico de Valpara\'iso (CCTVal),\\Universidad T\'ecnica Federico
    Santa Mar\'ia\\Casilla 110-V, Valpara\'iso, Chile}

  \address[2]{Department of Mathematical Sciences, Durham
    University,\\South Road, Durham DH1 3LE, United Kingdom}

  \address[3]{Via del Grano 33,\\
    Mogliano Veneto, Italy}
  
  \begin{abstract}
    The aim of this work is to present a series of concrete examples
    which illustrate how the computer algebra system \texttt{Cadabra}
    can be used to manipulate expressions appearing in General
    Relativity and other gravitational theories.
    We highlight the way in which \texttt{Cadabra}'s
    philosophy differs from other systems with related functionality.
    The use of various new built-in packages is discussed, and we show
    how such packages can also be created by end-users directly using
    the notebook interface.

    The current paper focuses on fairly generic applications in
    gravitational theories, including the use of differential forms,
    the derivation of field equations and the construction of their
    solutions.  A follow-up paper discusses more specific applications
    related to the analysis of \emph{gravitational waves}
    \cite{castillo-felisola20_cadab_python_algor_gener_relat_cosmol_ii}.
  \end{abstract}
\end{frontmatter}


\section{Introduction}
\label{sec:intro}

Algebraic manipulation of mathematical expressions is a common but
tedious part of most research in physics. Symbolic computer algebra
software has been used from the early days to help with this, and many
special-purpose systems have been built to deal with expression
manipulations specific to particular areas in physics. This is in
particular true for research in gravity; for a recent review of the
many uses of symbolic computer algebra in this field
see Ref.~\cite{MacCallum:2018csx}.

\texttt{Cadabra} is a relatively new, free and open-source standalone
computer algebra system,\footnote{The official site for \texttt{Cadabra}
  is \url{https://cadabra.science}, and the actual code is hosted at
  \url{https://github.com/kpeeters/cadabra2}.} which was designed
from the ground up to manipulate mathematical expressions which occur
in classical and quantum field theory~\cite{peeters07_cadab,
  peeters07_introd_cadab, peeters07_symbol_field_theor_with_cadab,
  Peeters:2018dyg}. In contrast to many special-purpose systems
written for sometimes very specific tasks, it aims to provide a wide
variety of basic field-theory building blocks, not only to tackle
gravity computations but also to provide support for things like
fermions and anti-commuting variables, algebra-valued objects,
component and abstract computations, tensor symmetries and various
others. Its main philosophy is to provide a simple-to-use `scratchpad'
for computations in field theory in its widest sense, to help with
computations which are too tedious to do by hand, while keeping them
close in form to what those computations would look like on paper. It
is programmable in Python, yet also accepts mathematical expressions
in standard \LaTeX{} notation. It has been used in a wide variety of
computations in high-energy physics and gravity, but also in different
fields such as nuclear physics~\cite{baran2019analytical}.

Because \texttt{Cadabra} tries to encourage and support a work-flow
which is close to how computations are done with pencil and paper, it
differs sometimes quite strongly from other computer algebra systems
with a wide scope.  In the present paper, and its followup
companion~\cite{castillo-felisola20_cadab_python_algor_gener_relat_cosmol_ii},
our goal is to show how gravity computations (which will at least in
spirit be familiar to many readers) can be done with
\texttt{Cadabra}. For a deeper look into using the system for advanced
gravity computations, see
e.g.~\cite{brewin19_using_cadab_tensor_comput_gener_relat}.

The approach of \verb=cadabra= is that (geometrical) objects are first
declared by assigning \verb=properties= to objects, after which they
can then be manipulated with \verb=algorithms=, which act according to
the previously assigned properties. A brief example is in order,
\begin{cadabra}
{a,b}::NonCommuting.
{a,b}::Distributable.
{b}::SelfAntiCommuting.
expr := b ( a + b + a b );
distribute(_);
\end{cadabra}
\begin{equation*}
  \begin{split}
    & b \left(a+b+a b\right)
    \\
    & b a+b a b
  \end{split}
\end{equation*}
In the above example \(a\) and \(b\) are the \emph{objects}, and we
assign the properties \verb=NonCommuting= and \verb=Distributable= to
them, and additionally assign the property \verb=SelfAntiCommuting= to
the object \(b\). The first assignment forbids the rearrangement of \(b a
b\) as \(a b^2\), while the last assignment ensures that \(b^2 =
0\). Notice that output is shown only when a command ends with a 
semi-colon (\verb=;=). 
\medskip

This paper is organised as follows. Section~\ref{sec:Formalism}
briefly introduces the basic concepts of General Relativity (see
Sec.~\ref{sec:GR-form}) and their implementation in \verb+cadabra+
(see Sec.~\ref{sec:header}). The code in Sec.~\ref{sec:header} is
intended to serve as a \emph{header} file, which can be called from
other \verb+cadabra+ notebooks in order to avoid the declaration of
``standard'' properties. In Sec.~\ref{sec:manip-tens} we exemplify the
manipulation of tensor expressions by writing down explicit
expressions for the Lanczos--Lovelock Lagrangians in Sec.~\ref{sec:LL},
and their field equations in Sec.~\ref{sec:LLfeq}. In
Sec.~\ref{sec:diff-forms} we explore the capabilities of
\verb+cadabra+ to manipulate differential forms. Specifically, we
obtain the Bianchi identities from the structural
equations.\footnote{The content of this section is available on the
  user contributed notebooks section of the official \texttt{cadabra}
  webpage. \label{foot:user-notebook}} Next, in
Sec.~\ref{sec:varprinc} the variational principle is exemplified by
extremising the Einstein--Hilbert
action.\textsuperscript{\ref{foot:user-notebook}} For reasons of
space, the variation of the Lanczos--Lovelock action is not addressed
in this paper. We then deal with the resolution of Einstein field
equations in Sec.~\ref{sec:sols}. In Sec.~\ref{sec:schw} we consider
the Schwarzschild spacetime, while Friedman--Robertson--Walker cases are
examined in Sec.~\ref{sec:frw}. Some conclusions are drawn in the
Sec.~\ref{sec:concl}. In \ref{sec:bench} we introduce a tool that
could help to improve the performance of calculations or long
routines.

\section{Formalism}
\label{sec:Formalism}


\subsection{Introduction to the formalism of General Relativity}
\label{sec:GR-form}

Let us start by giving a brief reminder of the ingredients of General
Relativity, both to set our conventions and to prepare for the
discussion of its properties formulated in \verb=Cadabra='s language.
General Relativity is currently the best model of gravitational
interactions, and was proposed in 1915 by
A.~Einstein~\cite{einstein15_zur_allgem_relat,einstein16_grund_allgem_relat},
as an attempt to conciliate the concepts introduced by the special
theory of relativity with those of gravitation. In his model, Einstein
proposed that the gravitational interaction is an effect of the
curvature of the spacetime. Meanwhile, the matter distribution
determines how the spacetime curves. This is sometimes called the
geometrisation of gravity.

Since the theory has to be invariant under general coordinate
transformations, its building blocks are \emph{tensors} (or more
generally \emph{tensor densities}). In General Relativity the
spacetime is assumed to be a pseudo-Riemannian manifold, whose
geometry is completely characterised by the metric tensor,
\(g_{\mu\nu}\). In order for the derivative of a tensor to be a
tensor, the concept of connection (\(\Gamma^{\lambda}{}_{\mu\nu}\))
has to be introduced, allowing to define a \emph{covariant derivative}
(\(\nabla_\mu = \partial_\mu + \Gamma^\bullet{}_{\mu \bullet}\)). The
condition of metricity, i.e. \(\nabla g = 0\), relates the
connection (the Levi-Civita connection) to the metric and partial
derivatives of it,
\begin{equation}
  \Gamma^{\mu}{}_{\nu\tau} = \frac{1}{2} g^{\mu\sigma} (
  \partial_{\tau}{g_{\nu\sigma}} + \partial_{\nu}{g_{\tau\sigma}} - \partial_{\sigma}{g_{\nu\tau}} ).
  \label{eq:def-chr}
\end{equation}
The Levi-Civita connection is symmetric in its lower indices,
\(\Gamma^{\mu}{}_{\nu\tau} = \Gamma^{\mu}{}_{\tau\nu}\), this property
is referred as \emph{torsion-free condition}.

The action of the commutator of covariant derivatives on a vector
yields an algebraic operator, dubbed the \emph{curvature tensor},
\begin{equation*}
  [\nabla_\mu , \nabla_\nu] V^\tau = R^{\tau}{}_{\sigma\mu\nu} V^\sigma,
\end{equation*}
where
\begin{equation}
  R^{\tau}{}_{\sigma\mu\nu} = \partial_{\mu}{\Gamma^{\tau}{}_{\nu\sigma}}
  - \partial_{\nu}{\Gamma^{\tau}{}_{\mu\sigma}}
  + \Gamma^{\tau}{}_{\mu\lambda} \Gamma^{\lambda}{}_{\nu\sigma}
  - \Gamma^{\tau}{}_{\nu\lambda} \Gamma^{\lambda}{}_{\mu\sigma}.
  \label{eq:def-curv}
\end{equation}
The curvature tensor, also known as the Riemann tensor, is
skew-symmetric in the last two indices, and additionally satisfies the
algebraic and differential Bianchi identities,
\begin{align*}
  R^{\tau}{}_{\sigma\mu\nu} + R^{\tau}{}_{\mu\nu\sigma} +
  R^{\tau}{}_{\nu\sigma\mu} & = 0,
  \\
  \nabla_\lambda R^{\tau}{}_{\sigma\mu\nu} + \nabla_\mu
  R^{\tau}{}_{\sigma\nu\lambda} + \nabla_\nu R^{\tau}{}_{\sigma\lambda\mu}
  & = 0.
\end{align*}
The contraction of the Riemann tensor are interesting geometrical
quantities,
\begin{equation}
  R_{\sigma\nu}  = R^{\tau}{}_{\sigma\tau\nu}
  \quad
  \text{and}
  \quad
  R = g^{\mu \nu} R_{\mu \nu},
\end{equation}
called the Ricci tensor and Ricci scalar (curvature) respectively.

From a physics perspective, the relevant geometrical object is the
\emph{Einstein tensor},
\begin{equation}
  G_{\mu \nu} = R_{\mu \nu} - \frac{1}{2} g_{\mu \nu} R.
  \label{eq:def-Eins}
\end{equation}
The field equations of General Relativity are obtained by extremising the
Einstein--Hilbert action,
\begin{equation}
  S = \frac{1}{2 \kappa} \int \mathrm{d}^4 x \, \sqrt{-g} \left( R - 2 \Lambda \right),
  \label{eq:EH-action}
\end{equation}
where \(\kappa\) is the coupling constant of gravity (inversely
proportional to the gravitational Newton constant \(G_N\)),
\(\Lambda\) is the cosmological constant, and the symbol \(g\) stands
for the determinant of the metric tensor.\footnote{The minus sign in
  front of the determinant is necessary because the spacetime has a
  Lorentzian signature.}

The interaction between matter and gravity is (formally) achieved
through the minimal coupling mechanism, i.e. starting from the action
on a flat spacetime and replacing the partial derivative by covariant
derivative, the Minkowski metric by the curved metric, and the flat
volume measure \(\mathrm{d}^4x\) by the invariant volume measure
\(\mathrm{d}^4 x \sqrt{-g}\). Hence, to the action in
Eq.~\eqref{eq:EH-action} one adds the matter action,
\begin{equation}
  S_{\text{mat}} = \int \mathrm{d}^4x \sqrt{-g}
  \Lag_{\text{mat}}(\psi, g, \nabla\psi).
  \label{eq:matter-action}
\end{equation}
This then leads to Einstein's equations, which set the Einstein tensor
$G_{\mu\nu}$ proportional to the
energy-momentum tensor, \(T_{\mu\nu}\), which encodes the properties
of the matter distribution.

\subsection{Introduction to the formalism of Cadabra: The header.cnb library}
\label{sec:header}

\begin{figure}[ht]
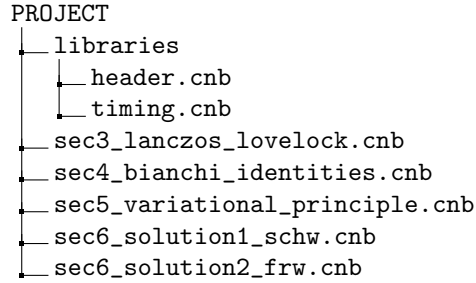

  \centering
  \begin{minipage}{7cm}
    \dirtree{%
.1 PROJECT.
.2 libraries.
.3 header.cnb.
.3 timing.cnb.
.2 sec3_lanczos_lovelock.cnb.
.2 sec4_bianchi_identities.cnb.
.2 sec5_variational_principle.cnb.
.2 sec6_solution1_schw.cnb.
.2 sec6_solution2_frw.cnb.
 }
  \end{minipage}
  \caption{\label{fig:project_structure} Organisation of notebooks
    discussed in the present paper.}
\end{figure}

The code presented in this article is organised as a project
containing a notebook for each chapter, as well as a separate library
with two \texttt{Cadabra} packages which can be re-used in other
computations.\footnote{The project can be downloaded
  from it official GitLab repository~\url{https://gitlab.com/cdbgr/cadabra-gravity-I}.}
The structure of the project is depicted in
figure~\ref{fig:project_structure}.

The \textit{header.cnb} notebook is included at the start of some of
the notebooks and defines a set of objects and related properties
which will be used throughout the discussion of tensor perturbations
in General Relativity. Providing this structure also allows us to
define a useful workflow for the follow up to this paper
\cite{castillo-felisola20_cadab_python_algor_gener_relat_cosmol_ii}.
The purpose of this is to avoid repetitive declarations and ensure
consistency between the notebooks. As expected, the file begins by
importing the global dependencies:
\StartLineAt{1}
\begin{cadabra}
import sympy
import cdb.core.manip as manip
import cdb.core.component as comp
import cdb.sympy.solvers as solv
\end{cadabra}
The last three imports are from the \texttt{Cadabra} standard
library~\cite{cadabra_man}, and provide common operations. The
\texttt{cdb.core.comp} library is useful for component calculations
and \texttt{cdb.sympy.solvers} is a simple \texttt{Cadabra} wrapper
for the equation solvers provided in the \texttt{sympy} library.

Although these imports appear to be regular Python packages, they are
in fact \texttt{Cadabra} notebooks and can be found at
\verb|<cadabra-root-dir>/lib/python3.x/site-packages/cdb|. When
\texttt{Cadabra} finds an \texttt{import} statement, it will not only
do a standard search in \texttt{sys.path} for Python packages, but it
will also search for notebook files, which are automatically converted
into Python scripts and imported using the native functionality.  Not
only does this make writing packages for \texttt{Cadabra} very
natural, but it also makes these packages very easy to read, as the
documentation is written next to the code in \LaTeX{} cells using the
\verb|\algorithm| command, and test code can be written under the
exported functions in ghost cells which are ignored when imported,
similarly to how \texttt{if __name__ == "__main__"} statements are
used in Python.

After this, the \textit{header.cnb} file defines one main function
\texttt{init_properties}, which accepts a \texttt{coordinates}
parameter containing the range of coordinates used through the
notebooks and a \texttt{metrics} parameter with the names of the
metric tensors required, and uses this to inject appropriate property
declarations into the \texttt{Cadabra} kernel. It begins by declaring
the coordinates and indices used in the notebooks:
\ContinueLineNumber
\begin{cadabra}
def init_properties(*, coordinates, metrics=[$g_{\mu\nu}$], signature=-1):
  """
  Initialise common property declarations which are shared between the
  two papers
  """
  @(coordinates)::Coordinate.
  index_list := {\mu,\nu,\rho,\sigma,\alpha,\beta,\gamma,\tau,\chi,\psi,\lambda,\lambda#}.
  @(index_list)::Indices(position=independent, values=@(coordinates)).
  Integer(index_list, Ex(rf"1..{len(coordinates)}"))
\end{cadabra}

The use of the \textit{pull-in} syntax \texttt{@(...)} allows us to
use Cadabra expressions inside other expressions, similarly to how
curly brackets are used in Python strings to include other objects
(e.g. \verb|'This is some {}'.format('text'))|. As the
\texttt{::Property} syntax expects a Cadabra expression on the left
hand side, not a variable name, we use it here to declare properties
on these expressions which are not hard-coded. As well as assigning
the \texttt{Indices} property to our index list, we also assign the
\texttt{Integer} property which makes the number of coordinates
countable, allowing functions like \verb|eliminate_kronecker| which
makes the substitution $\delta_{\mu}^{\mu} \rightarrow D$ to work. One
final thing to note is the \verb+#+, which declares an infinite set of
labelled indices (i.e. \verb+\lambda1+, \verb+\lambda2+) which is
useful to ensure that spare indices are always available (useful when
running code in loops and when doing higher perturbative orders of a
computation, for which it is not always easy to estimate how many
dummy indices will be required).

The function then associates the relevant properties to the metrics
defined in the \texttt{metrics} parameter: 
\StartLineAt{15}
\begin{cadabra}
  sig = Ex(signature)
  for metric in metrics:
    # Both lower indices: Metric
    for index in metric.top().indices():
      index.parent_rel = parent_rel_t.sub
    @(metric)::Metric(signature=@(sig)).

    # Both upper indices: InverseMetric
    for index in metric.top().indices():
      index.parent_rel = parent_rel_t.super
    @(metric)::InverseMetric(signature=@(sig)).

    # Mixed indices: KroneckerDelta
    for index in metric.top().indices():
      index.parent_rel = parent_rel_t.sub
      @(metric)::KroneckerDelta.
      index.parent_rel = parent_rel_t.super
\end{cadabra}
Here the \texttt{for} loops act over the two indices of the metric,
lowering them for the \texttt{Metric} declaration and raising them for
the \texttt{InverseMetric} declaration. Note that the $(1, 1)$ forms
of the metric tensor (e.g. $g^\mu{}_\nu$) must be declared as
\texttt{KroneckerDelta}. After this some standard symbols are defined 
\StartLineAt{33}
\begin{cadabra}
  # Symbols
  i::ImaginaryI.
  \delta{#}::KroneckerDelta.
  \partial{#}::PartialDerivative.
  \nabla{#}::Derivative.
  d{#}::Derivative.
\end{cadabra}

Finally, we introduce some of the standard GR objects. For each one,
multiple properties must be defined---we begin by assigning a
\texttt{Cadabra} identifier to each one and assigning a
\texttt{LaTeXForm} to it which controls how it will be rendered in
\LaTeX{}. The symmetries and dependencies of each are then defined; as
\texttt{Cadabra} makes very few assumptions about the objects one uses
any dependence on derivatives must be explicitly stated. 
\StartLineAt{40}
\begin{cadabra}
  # LaTeX Typography
  ch{#}::LaTeXForm("\Gamma").
  {rm{#},rc{#},sc}::LaTeXForm("R").
  ei{#}::LaTeXForm("G").
  Lmb{#}::LaTeXForm("\Lambda").

  # Symmetries
  ch^{\rho}_{\mu\nu}::TableauSymmetry(shape={2},indices={1,2}).
  rm^{\rho}_{\sigma\mu\nu}::TableauSymmetry(shape={1,1}, indices={2,3}).
  rc_{\mu\nu}::Symmetric.
  ei_{\mu\nu}::Symmetric.
\end{cadabra}

This completes the \texttt{init_properties} function, and the
remainder of the \texttt{header.cnb} file are some algebraic
definitions of the GR objects that are used throughout the notebooks,
allowing them to be substituted for each other. One can check the
sanity of the definition with a few testing lines,
\begin{cadabra}[numbers=none]
init_properties(coordinates=$t,x,y,z$, metrics=[$g_{\mu\nu}$, $\eta_{\mu\nu}$])
# Test that properties are declared correctly
assert eliminate_metric($g_{\mu\rho}x^{\rho\nu}$) == $x_{\mu}^{\nu}$
assert eliminate_kronecker($g_{\mu}^{\rho}x_{\rho\nu}$) == $x_{\mu\nu}$
assert eliminate_kronecker($\eta^{\mu}_{\mu}$) == $4$
assert eliminate_kronecker($g_{\mu}^{\mu}$) == $4$
\end{cadabra}

As some of them depend on definitions of symbols as derivatives in
order for index consistency to be maintained, they are defined as
functions returning expression objects so that they are only parsed
when called (after \texttt{init_properties} has been invoked) and not
during import where they would raise a parsing error
\StartLineAt{58}
\begin{cadabra}
def ch():
  return $ch^{\mu}_{\nu\tau} = \frac{1}{2} g^{\mu\sigma} (\partial_{\tau}{g_{\nu\sigma}} + \partial_{\nu}{g_{\tau\sigma}} - \partial_{\sigma}{g_{\nu\tau}})$
def rm():
  return $rm^{\tau}_{\sigma\mu\nu} = \partial_{\mu}{ch^{\tau}_{\nu\sigma}} - \partial_{\nu}{ch^{\tau}_{\mu\sigma}} + ch^{\tau}_{\mu\lambda} ch^{\lambda}_{\nu\sigma} - ch^{\tau}_{\nu\lambda} ch^{\lambda}_{\mu\sigma}$
def rc():
  return $rc_{\sigma\nu} = rm^{\tau}_{\sigma\tau\nu}$
def rs():
  return $sc = g^{\mu\nu} rc_{\mu\nu}$
def ei():
  return $ei_{\mu\nu} = rc_{\mu\nu} - \frac{1}{2} g_{\mu\nu} sc$
\end{cadabra}
These correspond to the relations/definitions
\begin{dgroup}
  \begin{dmath}
    \Gamma^{\mu}\,_{\nu \rho} = \tfrac{1}{2}g^{\mu \sigma}(\partial_{\rho}{g_{\nu \sigma}}+\partial_{\nu}{g_{\rho \sigma}}-\partial_{\sigma}{g_{\nu \rho}}),
  \end{dmath}
  \begin{dmath}
    R^{\rho}\,_{\sigma \mu \nu} = \partial_{\mu}{\Gamma^{\rho}\,_{\nu \sigma}}-\partial_{\nu}{\Gamma^{\rho}\,_{\mu \sigma}}+\Gamma^{\rho}\,_{\mu \lambda} \Gamma^{\lambda}\,_{\nu \sigma}-\Gamma^{\rho}\,_{\nu \lambda} \Gamma^{\lambda}\,_{\mu \sigma},
  \end{dmath}
  \begin{dmath}
    R_{\sigma \nu} = R^{\rho}\,_{\sigma \rho \nu},
  \end{dmath}
  \begin{dmath}
    R = g^{\mu \nu} R_{\mu \nu},
  \end{dmath}
  \begin{dmath}
    G_{\mu \nu} = R_{\mu \nu} - \tfrac{1}{2}g_{\mu \nu} R.
  \end{dmath}
\end{dgroup}

\section{Manipulation of tensorial expressions}
\label{sec:manip-tens}

\subsection{Explicit form of the Lanczos--Lovelock Lagrangian}
\label{sec:LL}

The Lanczos--Lovelock
Lagrangians~\cite{lanczos38_remar_proper_rieman_chris_tensor_four,lovelock69_uniquen_einst_field_equat_four_dimen_space}
are build in arbitrary dimensions to satisfy the same requirements as
the Einstein--Hilbert Lagrangian in General Relativity: (i) Invariance
under general coordinate transformations; (ii) local Lorentz symmetry,
and; (iii) Provide field equations that are second order partial
differential equations.

In four dimensions, the Lanczos--Lovelock Lagrangian is that of
Einstein--Hilbert (with cosmological constant) with the addition of a
topological terms, the Gau\ss--Bonnet
Lagrangian~\cite{lovelock69_uniquen_einst_field_equat_four_dimen_space}.
The generic Lanczos--Lovelock Lagrangian density in \(D\) dimensions
is given by a 
sum~\cite{lovelock71_einst_tensor_its,padmanabhan.13_lancz_lovel_model_gravit}
\begin{equation}
  \Lag = \sum_{p=0}^{n} \, a_p \, \Lag^{(D,p)},
\end{equation}
with \(n = \left[ \frac{D}{2} \right]\), and 
\begin{equation}
  \label{eq:cad:LLp}
  \Lag^{(D,p)} = \sqrt{g} \, \frac{1}{2^p} \, \delta_{a_1 \cdots a_{2p}}^{b_1 \cdots b_{2p}} \, R^{a_1 a_2}{}_{b_1 b_2} \cdots R^{a_{2p-1} a_{2p}}{}_{b_{2p-1} b_{2p}}.
\end{equation}

The purpose of this example is to find the form of the elements of the
series, i.e. Eq.~\eqref{eq:cad:LLp} after eliminating the generalised
delta. 

\subsubsection*{Notation and definitions}

In this example the position of the indices (unlike in General
Relativity) carry no meaning, and we shall write all of them as upper
indices. Hence, for this example the default behaviour of indices will
be enough. In this example we do not use the \verb+header.cnb+ file
from Sec.~\ref{sec:header}, consequently we shall explain some of the
declarations as they appear.

We start by declaring the indices, Kronecker delta, Levi-Civita
epsilon, and the partial derivative.
\StartLineAt{1}
\begin{cadabra}
{a#,b#,s#,t#,m,n}::Indices.
{a#,b#,s#,t#,m,n}::Integer(1..D).
\delta{#}::KroneckerDelta.
\epsilon{#}::EpsilonTensor(delta=-\delta).
\end{cadabra}
The hash symbol after the \texttt{a}, \texttt{b}, \texttt{s} and
\texttt{t} indices denotes that either of them followed by a number is
a valid index. The \emph{delta} argument of the Levi-Civita symbol
(\verb+EpsilonTensor+) serves to declare the signature of the
metric. In the example above, the value \verb+-\delta+ implies that
the---undeclared---metric is Lorentzian.

Next, we declare the the dependencies and symmetries of the curvature
tensors. In addition to the \texttt{TableauSymmetry} property
\texttt{cadabra} has the \texttt{Symmetric} and \texttt{AntiSymmetric}
properties, which endow the symmetry to all of the indices of the
object. 
\ContinueLineNumber
\begin{cadabra}
R^{s1 s2 s3 s4}::TableauSymmetry( shape={2,2}, indices={0,2,1,3} ).
R^{s1 s2}::Symmetric.
\end{cadabra}

In order to simplify the final expressions, we define a set of
substitution rules for the contractions of the curvature tensor,
\ContinueLineNumber
\begin{cadabra}
toR := {R^{s1 s2 s1 s2} = R, R^{s1 s2 s2 s1} = - R};
toRic := {R^{s1 s2 s1 s3} = R^{s2 s3}, 
  R^{s2 s1 s3 s1} = R^{s2 s3},
  R^{s1 s2 s3 s1} = - R^{s2 s3},
  R^{s2 s1 s1 s3} = - R^{s2 s3}};
\end{cadabra}

\subsubsection*{Zeroth order in curvature}

From the Eq.~\eqref{eq:cad:LLp}, it is obvious that the zeroth curvature
term is the cosmological constant monomial in the action. Thus, no
further work is needed.

\subsubsection*{Linear curvature Lagrangian}

From the Eq.~\eqref{eq:cad:LLp}, the linear curvature Lagrangian is
\begin{equation*}
  \Lag^{(D,1)} = \sqrt{g} \, \frac{1}{2} \, \delta_{a_1 a_{2}}^{b_1 b_{2}} \, R^{a_1 a_2}{}_{b_1 b_2}.
\end{equation*}
Since the measure (\(\sqrt{g}\)) does not carry indices, let us focus
on the remaining expression. The general strategy is to expand the
\emph{generalised} Kronecker delta, consider all possible
contractions, and use the substitution rules in order to present the
result in a simpler form. However, we need a few extra manipulations
in the middle as shown below.
\ContinueLineNumber
\begin{cadabra}
LL1 := 2/2 R^{a1 a2 b1 b2} \delta^{a1 b1 a2 b2};
expand_delta(LL1)
distribute(LL1)
eliminate_kronecker(LL1)
canonicalise(LL1)
rename_dummies(LL1)
substitute(LL1, toR)
substitute(LL1, toRic)
sort_product(LL1)
sort_sum(LL1)
canonicalise(LL1)
rename_dummies(LL1)
collect_factors(LL1);
\end{cadabra}
\begin{equation}
  \begin{split}
    & R^{{a_{1}} {a_{2}} {b_{1}} {b_{2}}} \delta^{{a_{1}} {b_{1}} {a_{2}} {b_{2}}}
    \\
    & R
  \end{split}
  \label{eq:LL1}
\end{equation}
In the definition of the expression \verb+LL1+, the factor \(2\) in
the numerator is to compensate a \((2p)!\) factor in the definition of
the generalised delta in \verb+cadabra+. The result of
Eq.~\eqref{eq:LL1} is the scalar curvature, which corresponds to the
Einstein--Hilbert Lagrangian.

Before moving toward terms with higher order in curvature, let us
introduce the programming capability of \verb+cadabra+. The
manipulations above, can be turned in to a python-like function, which
contain the set of algorithms to be applied to an input
expression. Hence, the above code will be replaced by,
\StartLineAt{12}
\begin{cadabra}
def LLmanip(ex):
  expand_delta(ex)
  distribute(ex)
  eliminate_kronecker(ex)
  canonicalise(ex)
  rename_dummies(ex)
  substitute(ex, toR)
  substitute(ex, toRic)
  sort_product(ex)
  sort_sum(ex)
  canonicalise(ex)
  rename_dummies(ex)
  collect_factors(ex)
  return(ex)
\end{cadabra}
and finally we define the linear curvature term of the
Lanczos--Lovelock series, and apply the newly define algorithm,
\ContinueLineNumber
\begin{cadabra}
LL1 := 2/2 R^{a1 a2 b1 b2} \delta^{a1 b1 a2 b2};
LLmanip(LL1);
\end{cadabra}
\begin{equation}
  \begin{split}
    & R^{{a_{1}} {a_{2}} {b_{1}} {b_{2}}} \delta^{{a_{1}} {b_{1}} {a_{2}} {b_{2}}}
    \\
    & R
  \end{split}
\end{equation}

\subsubsection*{Quadratic curvature Lagrangian}

From the Eq.~\eqref{eq:cad:LLp}, the quadratic curvature Lagrangian is
\begin{equation*}
  \Lag^{(D,2)} = \sqrt{g} \, \frac{1}{2^2} \, \delta_{a_1 a_2 a_3 a_4}^{b_1 b_2 b_3 b_4} \, R^{a_1 a_2}{}_{b_1 b_2} R^{a_3 a_4}{}_{b_3 b_4}.
\end{equation*}
We proceed as in the previous case,
\begin{cadabra}
LL2 := 4*3*2/2/2 R^{a1 a2 b1 b2} R^{a3 a4 b3 b4} \delta^{a1 b1 a2 b2 a3 b3 a4 b4};
LLmanip(LL2);
\end{cadabra}
\begin{dgroup*}
  \begin{dmath*}
    {}6R^{{a_{1}} {a_{2}} {b_{1}} {b_{2}}} R^{{a_{3}} {a_{4}} {b_{3}} {b_{4}}} \delta^{{a_{1}} {b_{1}} {a_{2}} {b_{2}} {a_{3}} {b_{3}} {a_{4}} {b_{4}}}
  \end{dmath*}
  \begin{dmath}{}{R}^{2}-4R^{{a_{1}} {a_{2}}} R^{{a_{1}} {a_{2}}}+R^{{a_{1}} {a_{2}} {a_{3}} {a_{4}}} R^{{a_{1}} {a_{2}} {a_{3}} {a_{4}}}
    \label{eq:LL2}
  \end{dmath}
\end{dgroup*}

The result of Eq.~\eqref{eq:LL2} is the scalar combination quadratic
in the curvature tensor known as Gau\ss--Bonnet term. Although in four
dimensions this is a topological term, in dimensions higher than four
generates dynamical field equations.

\subsubsection*{Cubic curvature Lagrangian}

From the Eq.~\eqref{eq:cad:LLp}, the cubic curvature Lagrangian is
\begin{equation*}
  \Lag^{(D,3)} = \sqrt{g} \, \frac{1}{2^3} \, \delta_{a_1 a_2 a_3 a_4 a_5 a_6}^{b_1 b_2 b_3 b_4 b_5 b_6} \, R^{a_1 a_2}{}_{b_1 b_2} R^{a_3 a_4}{}_{b_3 b_4} R^{a_5 a_6}{}_{b_5 b_6}.
\end{equation*}
And repeating the procedure above we get,
\ContinueLineNumber
\begin{cadabra}
LL3:= 6*5*4*3*2/2/2/2 R^{a1 a2 b1 b2} R^{a3 a4 b3 b4} R^{a5 a6 b5 b6} \delta^{a1 b1 a2 b2 a3 b3 a4 b4 a5 b5 a6 b6};
LLmanip(LL3);
\end{cadabra}
\begin{dgroup*}
  \begin{dmath*}
    {}90R^{{a_{1}} {a_{2}} {b_{1}} {b_{2}}} R^{{a_{3}} {a_{4}} {b_{3}} {b_{4}}} R^{{a_{5}} {a_{6}} {b_{5}} {b_{6}}} \delta^{{a_{1}} {b_{1}} {a_{2}} {b_{2}} {a_{3}} {b_{3}} {a_{4}} {b_{4}} {a_{5}} {b_{5}} {a_{6}} {b_{6}}}
  \end{dmath*}
  \begin{dmath}
    {}{R}^{3}-12R R^{{a_{1}} {a_{2}}} R^{{a_{1}} {a_{2}}}+3R R^{{a_{1}} {a_{2}} {a_{3}} {a_{4}}} R^{{a_{1}} {a_{2}} {a_{3}} {a_{4}}}+16R^{{a_{1}} {a_{2}}} R^{{a_{1}} {a_{3}}} R^{{a_{2}} {a_{3}}}+24R^{{a_{1}} {a_{2}}} R^{{a_{3}} {a_{4}}} R^{{a_{1}} {a_{3}} {a_{2}} {a_{4}}}-24R^{{a_{1}} {a_{2}}} R^{{a_{1}} {a_{3}} {a_{4}} {a_{5}}} R^{{a_{2}} {a_{3}} {a_{4}} {a_{5}}}+2R^{{a_{1}} {a_{2}} {a_{3}} {a_{4}}} R^{{a_{1}} {a_{2}} {a_{5}} {a_{6}}} R^{{a_{3}} {a_{4}} {a_{5}} {a_{6}}}-8R^{{a_{1}} {a_{2}} {a_{3}} {a_{4}}} R^{{a_{1}} {a_{5}} {a_{3}} {a_{6}}} R^{{a_{2}} {a_{6}} {a_{4}} {a_{5}}}
    \label{eq:LL3}
  \end{dmath}
\end{dgroup*}
This Lagrangian is known as the M\"uller-Hoissen term
\cite{mueller-hoissen85_spont_compac_with_quadr_cubic_curvat_terms}.


In \ref{sec:LL4} we present the Lanczos--Lovelock Lagrangian of order
four in curvature.

\subsection{Field equations of the Lanczos--Lovelock Lagrangians}
\label{sec:LLfeq}

The field equations derived from the Lanczos--Lovelock Lagrangians are
\cite{mueller-hoissen85_spont_compac_with_quadr_cubic_curvat_terms,mueller-hoissen90_from_chern_simon_to_gauss_bonnet,verwimp89_higher_dimen_gravit}
\begin{dmath}
  {G^{(p)}}_m^n = - \frac{1}{2^{p+1}} \, \delta_{m a_1 \cdots a_{2p}}^{n b_1 \cdots b_{2p}} \, R^{a_1 a_2}{}_{b_1 b_2} \cdots R^{a_{2p-1} a_{2p}}{}_{b_{2p-1} b_{2p}}
  = - \frac{(2p+1)!}{2^{p+1}} \, \delta_{[m}^n R^{a_1 a_2}{}_{a_1 a_2} \cdots R^{a_{2p-1} a_{2p}}{}_{a_{2p-1} a_{2p}]}.
  \label{eq:LLfeq}
\end{dmath}
We can use the same function \verb+LLmanip+ to obtain the expressions
for the field equations. Note that in the output below, in order to
shrink the expressions, we have written the piece accompanying the
Kronecker \(\delta\) as a Lagrangian itself.

As before, the term independent of the curvature needs no calculation,
\begin{dmath*}
G^{(0)}{}_m^n = - \frac{1}{2} \delta_m^n.
\end{dmath*}

While the field equations for the linear, quadratic and cubic are
shown below,
\begin{cadabra}
feqLL1 := - 3*2/2/2 R^{a1 a2 b1 b2} \delta^{m n a1 b1 a2 b2};
LLmanip(feqLL1);
\end{cadabra}
\begin{dgroup*}
  \begin{dmath*}
    {} - \frac{3}{2}R^{{a_{1}} {a_{2}} {b_{1}} {b_{2}}} \delta^{m n {a_{1}} {b_{1}} {a_{2}} {b_{2}}}
  \end{dmath*}
  \begin{dmath}
    {} R^{m n} - \frac{1}{2}R \delta^{m n}
  \end{dmath}
\end{dgroup*}
\begin{cadabra}
feqLL2 := - 5*4*3*2/2/2/2 R^{a1 a2 b1 b2} R^{a3 a4 b3 b4} \delta^{m n a1 b1 a2 b2 a3 b3 a4 b4};
LLmanip(feqLL2)
factor_out(_, $\delta^{m n}$)
substitute(_, $@(LL2) -> L^{(D,2)}$);
\end{cadabra}
\begin{dgroup*}
  \begin{dmath*}
    {}-15R^{{a_{1}} {a_{2}} {b_{1}} {b_{2}}} R^{{a_{3}} {a_{4}} {b_{3}} {b_{4}}} \delta^{m n {a_{1}} {b_{1}} {a_{2}} {b_{2}} {a_{3}} {b_{3}} {a_{4}} {b_{4}}}
  \end{dmath*}
  \begin{dmath}
    {}2R R^{m n}-4R^{{a_{1}} {a_{2}}} R^{m {a_{1}} n {a_{2}}}-4R^{m {a_{1}}} R^{n {a_{1}}}+2R^{m {a_{1}} {a_{2}} {a_{3}}} R^{n {a_{1}} {a_{2}} {a_{3}}} - \frac{1}{2}\delta^{m n} L^{(D,2)}
  \end{dmath}
\end{dgroup*}
\begin{cadabra}
feqLL3 := - 7*6*5*4*3*2/2/2/2/2 R^{a1 a2 b1 b2} R^{a3 a4 b3 b4} R^{a5 a6 b5 b6} \delta^{m n a1 b1 a2 b2 a3 b3 a4 b4 a5 b5 a6 b6};
LLmanip(feqLL3)
factor_out(_, $\delta^{m n}$)
substitute(_, $@(LL3) -> L^{(D,3)}$);
\end{cadabra}
\begin{dgroup*}
  \begin{dmath*}
    {}-315R^{{a_{1}} {a_{2}} {b_{1}} {b_{2}}} R^{{a_{3}} {a_{4}} {b_{3}} {b_{4}}} R^{{a_{5}} {a_{6}} {b_{5}} {b_{6}}} \delta^{m n {a_{1}} {b_{1}} {a_{2}} {b_{2}} {a_{3}} {b_{3}} {a_{4}} {b_{4}} {a_{5}} {b_{5}} {a_{6}} {b_{6}}}
  \end{dmath*}
  \begin{dmath}
    {}3{R}^{2} R^{m n}-12R R^{{a_{1}} {a_{2}}} R^{m {a_{1}} n {a_{2}}}-12R R^{m {a_{1}}} R^{n {a_{1}}}+6R R^{m {a_{1}} {a_{2}} {a_{3}}} R^{n {a_{1}} {a_{2}} {a_{3}}}+24R^{n {a_{1}}} R^{{a_{2}} {a_{3}}} R^{m {a_{2}} {a_{1}} {a_{3}}}+24R^{{a_{1}} {a_{2}}} R^{{a_{1}} {a_{3}}} R^{m {a_{2}} n {a_{3}}}-12R^{{a_{1}} {a_{2}}} R^{m {a_{1}} {a_{3}} {a_{4}}} R^{n {a_{2}} {a_{3}} {a_{4}}}-12R^{m n} R^{{a_{1}} {a_{2}}} R^{{a_{1}} {a_{2}}}+24R^{m {a_{1}}} R^{n {a_{2}}} R^{{a_{1}} {a_{2}}}+24R^{m {a_{1}}} R^{{a_{2}} {a_{3}}} R^{n {a_{2}} {a_{1}} {a_{3}}}+24R^{{a_{1}} {a_{2}}} R^{m {a_{3}} n {a_{4}}} R^{{a_{1}} {a_{3}} {a_{2}} {a_{4}}}-24R^{{a_{1}} {a_{2}}} R^{m {a_{3}} {a_{1}} {a_{4}}} R^{n {a_{3}} {a_{2}} {a_{4}}}-12R^{m {a_{1}}} R^{n {a_{2}} {a_{3}} {a_{4}}} R^{{a_{1}} {a_{2}} {a_{3}} {a_{4}}}+3R^{m n} R^{{a_{1}} {a_{2}} {a_{3}} {a_{4}}} R^{{a_{1}} {a_{2}} {a_{3}} {a_{4}}}-12R^{n {a_{1}}} R^{m {a_{2}} {a_{3}} {a_{4}}} R^{{a_{1}} {a_{2}} {a_{3}} {a_{4}}}-12R^{m {a_{1}} n {a_{2}}} R^{{a_{1}} {a_{3}} {a_{4}} {a_{5}}} R^{{a_{2}} {a_{3}} {a_{4}} {a_{5}}}-24R^{m {a_{1}} {a_{2}} {a_{3}}} R^{n {a_{4}} {a_{2}} {a_{5}}} R^{{a_{1}} {a_{5}} {a_{3}} {a_{4}}}+6R^{m {a_{1}} {a_{2}} {a_{3}}} R^{n {a_{1}} {a_{4}} {a_{5}}} R^{{a_{2}} {a_{3}} {a_{4}} {a_{5}}} - \frac{1}{2}\delta^{m n} L^{(D,3)}
  \end{dmath}
\end{dgroup*}

\section{Language of differential forms}
\label{sec:diff-forms}


A very useful calculation tool in physics is the exterior differential
calculus, which---unlike tensor calculus---deals only with the set of
completely anti-symmetric (differentiable) tensors of
\((0,p)\)-type, for \(0 \leq p \leq D = \dim(M)\). For a given value
of \(p\), the elements are called \emph{differential forms of degree}
\(p\) or \(p\)-forms, and they form a \(\mathcal{F}(M)\)-module
denoted by \(\Omega^p(M)\).\footnote{Note that the zero-forms on \(M\)
are just functions on the manifold, and therefore \(\Omega^0(M) =
\mathcal{F}(M)\).} The generalisation of the tensor product to a
product that preserves the antisymmetry of the result is dubbed
\emph{wedge} product, \(\wedge: \Omega^p(M) \times \Omega^q(M) \to
\Omega^{p+q}(M)\), and provides structure of algebra to the space of
all exterior differential forms,
\begin{equation*}
\Omega(M) = \sum_{p=0}^D \Omega^p(M),
\end{equation*}
known as exterior algebra. One can endow the exterior algebra with
more structure by defining the \emph{exterior derivative}, a smooth
map \(\mathrm{d}: \Omega^p(M) \to \Omega^{p+1}(M)\) satisfying the
(super)Leibniz rule,
\begin{equation*}
\mathrm{d} (\alpha \wedge \beta) = \mathrm{d}\alpha \wedge \beta +
(-1)^p \alpha \wedge \mathrm{d}\beta \condition{ for \(\alpha \in
  \Omega^p(M)\) and \(\beta \in \Omega^q(M)\)},
\end{equation*}
and nilpotency, i.e. \(\mathrm{d}^2 = 0\), and the \emph{Hodge star
  operator}, a maps \(\star: \Omega^p(M) \to \Omega^{D-p}(M)\), which
allows to define a symmetric scalar product among \(p\)-forms,
\begin{equation*}
(\alpha,\beta) = \int \alpha \wedge \star \beta \condition{ for
  \(\alpha,\beta \in \Omega^p(M)\)}.
\end{equation*}

A clear advantage of using differential forms over tensors is that
there are only a finite set of independent forms. However, in order to
encode the information within symmetric or mixed tensors, one have to
allow differential forms to be valued on a Lie algebra.\footnote{For
  gauge theories this algebra corresponds to the algebra of the gauge
  group, while for General Relativity it is the algebra of the Lorentz
  group. It is worth mentioning that there are modified theories of
  gravity which consider extended algebras of the gravitational
  sector, such as the de Sitter, Anti de Sitter or Poincar\'e
  algebras.} Hence, in the following we consider that the Lie algebra
behind the gravitational theory is the Lorentz algebra.

In this formalism, the geometric information carried by the metric and
affine connection is encoded in the vielbein 1-form (\(e^a\)) and the spin
connection \(1\)-form (\(\omega^{ab}\)). From them, one can calculate
the curvature and torsion \(2\)-forms using the structural equations
of Cartan,
\begin{equation*}
  \begin{split}
    \mathrm{d}{e^{a}}+\omega^{a}\,_{b}\wedge e^{b} & = \mathrm{T}^{a},
    \\
    \mathrm{d}{\omega^{a}\,_{b}}+\omega^{a}\,_{m}\wedge
    \omega^{m}\,_{b} & =\mathrm{R}^{a}\,_{b}.
  \end{split}
\end{equation*}

\subsection{Bianchi identities from Cartan structural equations}
\label{sec:str_eq}

\subsubsection*{Definitions}

As in previous examples, for this example we won't use the
declarations in the \verb!header.cnb! file. Hence we need to declare
the indices, derivative (in this case the exterior derivative), and
the geometrical objects---i.e. the differential forms.
\StartLineAt{1}
\begin{cadabra}
{a,b,c,l,m,n}::Indices.
d{#}::ExteriorDerivative;.
d{#}::LaTeXForm("\mathrm{d}").
T{#}::LaTeXForm("\mathrm{T}").
R{#}::LaTeXForm("\mathrm{R}").
{e^{a}, \omega^{a}_{b}}::DifferentialForm(degree=1); 
{T^{a}, R^{a}_{b}}::DifferentialForm(degree=2);
\end{cadabra}
In the above code, we have introduced the property \verb+DifferentialForm+,
which through its argument \verb+degree=p+ assigns the property
\(p\)-form to a given object or set of them, and
\verb+ExteriorDerivative+ which defines the exterior derivative to the
symbol \texttt{d}.

\subsubsection*{Cartan structural equations}

Now, we define the structural equations.
\ContinueLineNumber
\begin{cadabra}
struc1 := d{e^{a}} + \omega^{a}_{b} ^ e^{b} - T^{a} = 0;
struc2 := d{\omega^{a}_{b}} + \omega^{a}_{m} ^ \omega^{m}_{b} - R^{a}_{b} = 0;
\end{cadabra}
\begin{equation*}
  \begin{split}
    & \mathrm{d}{e^{a}}+\omega^{a}\,_{b}\wedge e^{b}-\mathrm{T}^{a} = 0
    \\
    & \mathrm{d}{\omega^{a}\,_{b}}+\omega^{a}\,_{m}\wedge \omega^{m}\,_{b}-\mathrm{R}^{a}\,_{b} = 0
  \end{split}
\end{equation*}

In the following, we will also use the structural equations as
definitions of the exterior derivatives of the vielbein and spin
connection 1-forms. Therefore, we shall utilise the \texttt{isolate}
algorithm---from the \texttt{cdb.core.manip} library---to define
substitution rules.
\begin{cadabra}
from cdb.core.manip import isolate 
de:= @(struc1):
isolate(de, $d{e^{a}}$);
\end{cadabra}
\begin{dmath*}
  \mathrm{d}{e^{a}} = -\omega^{a}\,_{b}\wedge e^{b}+\mathrm{T}^{a}
\end{dmath*}
\begin{cadabra}
domega := @(struc2):
isolate(domega, $d{\omega^{a}_{b}}$);
\end{cadabra}
\begin{dmath*}
  \mathrm{d}{\omega^{a}\,_{b}} = -\omega^{a}\,_{m}\wedge \omega^{m}\,_{b}+\mathrm{R}^{a}\,_{b}
\end{dmath*}

\subsubsection*{Bianchi identities}

The bianchi identities are obtained by applying the exterior
derivative to the structural equations.

\paragraph*{First Bianchi identity}

We apply the exterior derivative to the expression \verb+struc1+, and
then distribute and apply the Leibniz rule (\verb+product_rule+),
\begin{cadabra}
Bianchi1 :=  d{ @(struc1) };
distribute(Bianchi1)
product_rule(_);
\end{cadabra}
\begin{dgroup*}
  \begin{dmath*}
    \mathrm{d}\left(\mathrm{d}{e^{a}}+\omega^{a}\,_{b}\wedge
      e^{b}-\mathrm{T}^{a}\right) = 0
  \end{dmath*}
  \begin{dmath*}
    \mathrm{d}{\omega^{a}\,_{b}}\wedge e^{b}-\omega^{a}\,_{b}\wedge
    \mathrm{d}{e^{b}}-\mathrm{d}{\mathrm{T}^{a}} = 0
  \end{dmath*}
\end{dgroup*}

Now, the expression has the exterior derivative of the vielbein and
spin connection. Hence, we substitute the rules \verb+de+ and
\verb+domega+,
\begin{cadabra}
substitute(Bianchi1, de, repeat=True)
substitute(Bianchi1, domega, repeat=True)
distribute(_)
rename_dummies(_);
\end{cadabra}
\begin{dmath*}
  \mathrm{R}^{a}\,_{b}\wedge e^{b}-\omega^{a}\,_{b}\wedge \mathrm{T}^{b}-\mathrm{d}{\mathrm{T}^{a}} = 0
\end{dmath*}

In tensor form, the above relation is expressed as
\begin{equation*}
  R^{\lambda}{}_{[\rho\mu\nu]} - \nabla_{[\mu} T^{\lambda}{}_{\nu\rho]} - T^{\lambda}{}_{\sigma[\rho} T^{\sigma}{}_{\mu\nu]} = 0.
\end{equation*}
The above expression is the \emph{algebraic} Bianchi identity, which
in the absence of torsion takes the well-known form, 
\begin{equation*}
  R^{\lambda}{}_{[\rho\mu\nu]} = R^{\lambda}{}_{\mu\nu\rho} + R^{\lambda}{}_{\rho\mu\nu} + R^\lambda{}_{\nu\rho\mu} = 0.
\end{equation*}

\paragraph*{Second Bianchi identity}

Applying the exterior derivative to \verb+struc2+, we get
\begin{cadabra}
Bianchi2 := d{ @(struc2) };
distribute(Bianchi2)
product_rule(_);
\end{cadabra}
\begin{dgroup*}
  \begin{dmath*}
    \mathrm{d}\left(\mathrm{d}{\omega^{a}\,_{b}}+\omega^{a}\,_{m}\wedge
      \omega^{m}\,_{b}-\mathrm{R}^{a}\,_{b}\right) = 0
  \end{dmath*}
  \begin{dmath*}
    \mathrm{d}{\omega^{a}\,_{m}}\wedge
    \omega^{m}\,_{b}-\omega^{a}\,_{m}\wedge
    \mathrm{d}{\omega^{m}\,_{b}}-\mathrm{d}{\mathrm{R}^{a}\,_{b}} = 0
  \end{dmath*}
\end{dgroup*}

Here, only the exterior derivative of \(\omega\) appears, then
\begin{cadabra}
substitute(Bianchi2, domega, repeat=True)
distribute(_)
rename_dummies(_);
\end{cadabra}
\begin{dmath*}
  \mathrm{R}^{a}\,_{c}\wedge \omega^{c}\,_{b}-\omega^{a}\,_{c}\wedge
  \mathrm{R}^{c}\,_{b}-\mathrm{d}{\mathrm{R}^{a}\,_{b}} = 0
\end{dmath*}

The above result is the \emph{differential} Bianchi identity, that is
written in tensor form as
\begin{equation*}
  R^\lambda{}_{\sigma[\mu\nu;\rho]} + R^\lambda{}_{\sigma\tau[\rho} T^{\tau}{}_{\mu\nu]} = 0,
\end{equation*}
which in the absence of torsion takes the well-known form,
\begin{equation*}
  R^\lambda{}_{\sigma[\mu\nu;\rho]} = R^\lambda{}_{\sigma\mu\nu;\rho} + R^\lambda{}_{\sigma\nu\rho;\mu} + R^\lambda{}_{\sigma\rho\mu;\nu} = 0.
\end{equation*}

\section{Einstein equations from a variational principle}
\label{sec:varprinc}

In this example we derive the field equations of General Relativity
from Einstein--Hilbert action.

\subsection{Definitions}

For this example we shall import our \texttt{header} file from
Sec.~\ref{sec:header}. Note that despite the indices are declared as
four-dimensional, and the coordinates as spherical, the result is
valid for a generic choice of coordinates and dimension.
\StartLineAt{1}
\begin{cadabra}
from libraries.header import *
init_properties(coordinates=$t, r, \theta, \phi$)
\end{cadabra}

Then, we declare the \LaTeX{} output of a variety of expressions that
will show up in the process.
\ContinueLineNumber
\begin{cadabra}
Lm::LaTeXForm("\mathcal{L}_{\text{mat}}").
Dg::LaTeXForm("\sqrt{-g}").
dg{#}::LaTeXForm("\delta{g}").
dLm::LaTeXForm("\delta{\mathcal{L}_{\text{mat}}}").
dDg::LaTeXForm("\delta{\sqrt{-g}}").
dCn{#}::LaTeXForm("\delta{\Gamma}").
dR{#}::LaTeXForm("\delta{R}").
\end{cadabra}

\subsection{Variational problem}

Our starting point is the Einstein--Hilbert action, amended with a
cosmological constant term.
\begin{cadabra}
action := S = \int{Dg [ \frac{1}{2 \kappa} ( sc - 2 \Lambda) + Lm ]}{x}.
distribute(_);
\end{cadabra}
\begin{dmath*}
S = \int \left(\frac{1}{2} \sqrt{-g} {\kappa}^{-1} R-\sqrt{-g} {\kappa}^{-1} \Lambda+\sqrt{-g} \mathcal{L}_{\text{mat}}\right)\,\,{\rm d}x
\end{dmath*}
Next, we build the variation of the objects (fields) present in the
action.

The variation of the metric determinant is given by
\begin{cadabra}
deltaMetricDeterminant := dDg = - \frac{1}{2} Dg g_{\mu\nu} dg^{\mu\nu};
\end{cadabra}
\begin{dmath*}
\delta{\sqrt{-g}} =  - \frac{1}{2}\sqrt{-g} g_{\mu \nu} \delta g^{\mu \nu}
\end{dmath*}

By definition, the variation of the matter Lagrangian with respect to
the metric gives the stress tensor,
\begin{cadabra}
matterLagrangianVariation := Dg T_{\mu\nu} dg^{\mu\nu} = - 2 d{Dg Lm};
product_rule(_)
distribute(_)
substitute(_, $d{Dg} -> dDg$)
substitute(_, deltaMetricDeterminant)
manip.swap_sides(_);
\end{cadabra}
\begin{dgroup*}[noalign]
  \begin{dmath*}
    \sqrt{-g} T_{\mu \nu} \delta g^{\mu \nu} =
    -2\delta\left({\sqrt{-g} \mathcal{L}_{\text{mat}}}\right)
  \end{dmath*}
  \begin{dmath*}
    \sqrt{-g} g_{\mu \nu} \delta g^{\mu \nu}
    \mathcal{L}_{\text{mat}}-2\sqrt{-g}
    \delta{\mathcal{L}_{\text{mat}}} = \sqrt{-g} T_{\mu \nu} \delta
    g^{\mu \nu}
  \end{dmath*}
\end{dgroup*}

Next comes the variation of the Christoffel symbols. We then isolate
the derivative of the variation, as that object will appear in the
variation of the Riemann tensor.
\begin{cadabra}
deltaDerivativeCn := \nabla_{\sigma}{dCn^{\mu}_{\nu\tau}} = \partial_{\sigma}{dCn^{\mu}_{\nu\tau}}
  + ch^{\mu}_{\sigma\gamma} dCn^{\gamma}_{\nu\tau}
  - ch^{\gamma}_{\sigma\nu} dCn^{\mu}_{\gamma\tau}
  - ch^{\gamma}_{\sigma\tau} dCn^{\mu}_{\nu\gamma};
manip.isolate(_, $\partial_{\sigma}{dCn^{\mu}_{\nu\tau}}$);
\end{cadabra}
\begin{dgroup*}
  \begin{dmath*}
    \nabla_{\sigma}{\delta\Gamma^{\mu}\,_{\nu \tau}} = \partial_{\sigma}{\delta\Gamma^{\mu}\,_{\nu \tau}}+\Gamma^{\mu}\,_{\sigma \gamma} \delta\Gamma^{\gamma}\,_{\nu \tau}-\Gamma^{\gamma}\,_{\sigma \nu} \delta\Gamma^{\mu}\,_{\gamma \tau}-\Gamma^{\gamma}\,_{\sigma \tau} \delta\Gamma^{\mu}\,_{\nu \gamma}
  \end{dmath*}
  \begin{dmath*}
    \partial_{\sigma}{\delta\Gamma^{\mu}\,_{\nu \tau}} = -\Gamma^{\mu}\,_{\sigma \gamma} \delta\Gamma^{\gamma}\,_{\nu \tau}+\Gamma^{\gamma}\,_{\sigma \nu} \delta\Gamma^{\mu}\,_{\gamma \tau}+\Gamma^{\gamma}\,_{\sigma \tau} \delta\Gamma^{\mu}\,_{\nu \gamma}+\nabla_{\sigma}{\delta\Gamma^{\mu}\,_{\nu \tau}}
  \end{dmath*}
\end{dgroup*}
The variation of the Riemann tensor involves covariant derivatives of
the variation of the Christoffel connection.
\begin{cadabra}
deltaRiemannTensor = vary( rm(), $ch^{\mu}_{\nu\tau}->dCn^{\mu}_{\nu\tau}, rm^{\tau}_{\sigma\mu\nu} -> dR^{\tau}_{\sigma\mu\nu}$);
sort_product(_)
substitute(_,deltaDerivativeCn)
meld(_);
\end{cadabra}
\begin{dgroup*}
  \begin{dmath*}
    \delta R^{\tau}\,_{\sigma \mu \nu} =
    \partial_{\mu}{\delta\Gamma^{\tau}\,_{\nu
        \sigma}}-\partial_{\nu}{\delta\Gamma^{\tau}\,_{\mu
        \sigma}}+\delta\Gamma^{\tau}\,_{\mu \lambda}
    \Gamma^{\lambda}\,_{\nu \sigma}+\Gamma^{\tau}\,_{\mu \lambda}
    \delta\Gamma^{\lambda}\,_{\nu \sigma}-\delta\Gamma^{\tau}\,_{\nu
      \lambda} \Gamma^{\lambda}\,_{\mu \sigma}-\Gamma^{\tau}\,_{\nu
      \lambda} \delta\Gamma^{\lambda}\,_{\mu \sigma}
  \end{dmath*}
  \begin{dmath*}
    \delta R^{\tau}\,_{\sigma \mu \nu} =
    \nabla_{\mu}{\delta\Gamma^{\tau}\,_{\nu
        \sigma}}-\nabla_{\nu}{\delta\Gamma^{\tau}\,_{\mu \sigma}}
  \end{dmath*}
\end{dgroup*}

From the last expression one obtains the variation of the Ricci
tensor, dubbed \emph{Palatini identity}:
\begin{cadabra}
deltaRicciTensor = vary( rc(), $rm^{\tau}_{\sigma\tau\nu} -> dR^{\tau}_{\sigma\tau\nu}, rc_{\sigma\nu} -> dR_{\sigma\nu}$);
substitute(_, deltaRiemannTensor); 
\end{cadabra}
\begin{dmath*}
  \delta R_{\sigma \nu} = \nabla_{\tau}{\delta\Gamma^{\tau}\,_{\nu \sigma}}-\nabla_{\nu}{\delta\Gamma^{\tau}\,_{\tau \sigma}}
\end{dmath*}
Similarly, the variation of the scalar curvature yields
\begin{cadabra}
deltaScalarCurvature = vary( rs(), $sc-> dR, rc_{\sigma\nu} -> dR_{\sigma\nu}, g^{\mu\nu}->dg^{\mu\nu}$);
substitute(_, deltaRicciTensor)
distribute(_)
substitute(_, $\nabla_{\sigma}{dCn^{\mu}_{\nu\tau}} g^{\gamma\lambda} -> \nabla_{\sigma}{dCn^{\mu}_{\nu\tau} g^{\gamma\lambda}}$)
canonicalise(_);
\end{cadabra}
\begin{dmath*}
  \delta R = \nabla_{\nu}\left({\delta\Gamma^{\nu}\,_{\tau \sigma} g^{\tau \sigma}}\right)-\nabla_{\nu}\left({\delta\Gamma^{\tau}\,_{\tau \sigma} g^{\nu \sigma}}\right)+R_{\nu \sigma} \delta g^{\nu \sigma}
\end{dmath*}

We can now vary the action,
\begin{cadabra}
deltaAction = vary( $@(action):$, $Dg -> dDg, sc -> dR, Lm -> d{Lm}, S -> d{S}$);
substitute(_, deltaMetricDeterminant)
substitute(_, matterLagrangianVariation)
substitute(_, deltaScalarCurvature)
distribute(_)
rename_dummies(_)
factor_out(_, $dg^{\mu\nu}, Dg$);
\end{cadabra}
\begin{dgroup*}
  \begin{dmath*}
    \delta{S} = \int \left(\frac{1}{2}\delta{\sqrt{-g}} {\kappa}^{-1} R+\frac{1}{2}\sqrt{-g} {\kappa}^{-1} \delta R-\delta{\sqrt{-g}} {\kappa}^{-1} \Lambda+\delta{\sqrt{-g}} \mathcal{L}_{\text{mat}}+\sqrt{-g} \delta{\mathcal{L}_{\text{mat}}}\right)\,\,{\rm d}x
  \end{dmath*}
  \begin{dmath*}
    \delta{S} = \int \left(\sqrt{-g} \delta g^{\mu \nu} \left( - \frac{1}{4}g_{\mu \nu} {\kappa}^{-1} R+\frac{1}{2}{\kappa}^{-1} R_{\mu \nu}+\frac{1}{2}g_{\mu \nu} {\kappa}^{-1} \Lambda - \frac{1}{2}T_{\mu \nu}\right)+\sqrt{-g} \left(\frac{1}{2}{\kappa}^{-1} \nabla_{\mu}\left(\delta\Gamma^{\mu}\,_{\nu \tau} g^{\nu \tau}\right) - \frac{1}{2}{\kappa}^{-1} \nabla_{\tau}\left(\delta\Gamma^{\mu}\,_{\mu \nu} g^{\tau \nu}\right)\right)\right)\,\,{\rm d}x
  \end{dmath*}
\end{dgroup*}

The terms proportional to $\delta g^{\mu\nu}$ are the Einstein
equations, the rest is a total derivative.
In order to rewrite this term in the familiar form, let us first
assign to a variable the term in the first bracket on the right-hand
side. To select a piece of an expression one uses the array notation
of python: (i) the first level of the expression is a equality,
therefore the \texttt{[0]} and \texttt{[1]} components represent the
left-hand side and right-hand side respectively, thus we chose the
\texttt{[1]} component; (ii) the left-hand is an integration, and the
\texttt{[0]} and \texttt{[1]} components represent the argument and
integration variable respectively, thus we have to select the
\texttt{[0]} component; (iii) the argument of the integral is the sum
of two terms, since we are interested in the first, we have to select
the \texttt{[0]} component, and finally; (iv) from that term the
bracket is the \texttt{[2]} component. Therefore, the expression of
interest is \texttt{deltaAction[1][0][0][2]}. Hence,
\begin{cadabra}
t1 = deltaAction[1][0][0][2]
eom:= 2 \kappa @(t1) = 0;
distribute(_)
collect_factors(_)
manip.to_rhs(_, $\kappa T_{\mu\nu}$);
\end{cadabra}
\begin{dgroup*}[noalign]
  \begin{dmath*}
    2\kappa \left( - \frac{1}{4}g_{\mu \nu} {\kappa}^{-1} R+\frac{1}{2}{\kappa}^{-1} R_{\mu \nu}+\frac{1}{2}g_{\mu \nu} {\kappa}^{-1} \Lambda - \frac{1}{2}T_{\mu \nu}\right) = 0
  \end{dmath*}
  \begin{dmath*}
    - \frac{1}{2}g_{\mu \nu} R+R_{\mu \nu}+g_{\mu \nu} \Lambda = \kappa T_{\mu \nu}
  \end{dmath*}
\end{dgroup*}
These are the Einstein equations, which can be written in terms of the
Einstein tensor, \(G_{\mu\nu}\),
\begin{cadabra}
EinsteinEq := @(eom):
substitute(EinsteinEq, manip.swap_sides(ei()));
\end{cadabra}
\begin{dmath*}
  G_{\mu \nu}+g_{\mu \nu} \Lambda = \kappa T_{\mu \nu}
\end{dmath*}
or in the Ricci form, which is obtained after eliminating the scalar
curvature from the Einstein equations. Therefore, we first calculate
the trace to isolate the scalar curvature 
\begin{cadabra}
trEinsteinEq := @(eom):
manip.multiply_through(_,$g^{\mu\nu}$);
distribute(_)
substitute(_, $g^{\mu \nu} T_{\mu \nu} = T$)
substitute(_, manip.swap_sides(rs()))
eliminate_metric(_)
eliminate_kronecker(_)
manip.isolate(trEinsteinEq, $sc()$);
\end{cadabra}
\begin{dgroup*}[noalign]
  \begin{dmath*}
    g^{\mu \nu} \left( - \frac{1}{2}g_{\mu \nu} R+R_{\mu \nu}+g_{\mu \nu} \Lambda\right) = g^{\mu \nu} \kappa T_{\mu \nu}
  \end{dmath*}
  \begin{dmath*}
    R = -T \kappa+4\Lambda
  \end{dmath*}
\end{dgroup*}
and substitute into the field equations,
\begin{cadabra}
RicciFormEq := @(eom):
substitute(_, trEinsteinEq)
distribute(_);
manip.isolate(_, $rc_{\mu \nu}$)
factor_out(_, $\kappa$);
\end{cadabra}
\begin{dgroup*}[noalign]
  \begin{dmath*}{}\frac{1}{2}g_{\mu \nu} T \kappa-g_{\mu \nu} \Lambda+R_{\mu \nu} = \kappa T_{\mu \nu}
  \end{dmath*}
  \begin{dmath*}
    R_{\mu \nu} = g_{\mu \nu} \Lambda+\kappa \left(T_{\mu \nu} - \frac{1}{2}g_{\mu \nu} T\right)
  \end{dmath*}
\end{dgroup*}

Note that in the above manipulation we have used that \(\delta^\mu_\mu
= 4\),\footnote{The values of the indices have been set in the
  \texttt{header} file.} and consequently the field equations in the
form of Ricci in arbitrary dimension gets factor modifications, unlike
the standard Einstein's form.

\section{Spacetime solutions}
\label{sec:sols}

In this section we solve Einstein field equations for Schwarzschild
and Friedman--Robertson--Walker spacetimes. In both examples we upload
the definitions on our \texttt{header} file in Sec.~\ref{sec:header}.
Furthermore, for the Friedman--Robertson--Walker example, we define
the Einstein and energy-momentum tensors.

\subsubsection*{Definitions}

The code below will serve as common ground for both cases.
\StartLineAt{1}
\begin{cadabra}
from libraries.header import *
init_properties(coordinates=$t,r,\theta,\phi$, metrics=[$g_{\mu\nu}$, $\eta_{\mu\nu}$])
\end{cadabra}

\subsection{Schwarzschild spacetime}
\label{sec:schw}

In 1916, a few months after the publication of Einstein's equations,
the first nontrivial solution was found by Schwarzschild, who intended
to model the gravitational field of an isolated spherically symmetric
object.

The metric compatible with the three-dimensional spherical symmetry
has a line element given by
\begin{equation}
  \mathrm{d}s^2(g) = - \exp(A(t,r))  \mathrm{d}t^2 + \exp(B(t,r))
  \mathrm{d}r^2 + r^2 \mathrm{d}\Omega_{(2)}^2,
  \label{eq:schw-ansatz}
\end{equation}
where \(\mathrm{d}\Omega_{(2)}^2\) is the line element of a
two-dimensional sphere, i.e.
\begin{equation*}
\mathrm{d}\Omega_{(2)}^2 = \mathrm{d}\theta^2 + \sin^2 (\theta) \mathrm{d}\phi^2.
\end{equation*}

The above metric will be the starting point of our calculation, with
the exception that for simplicity, we shall consider the functions $A$
and $B$ as time-independent. Such condition could be derived from the
field equations, and is a corollary of the dubbed \emph{Birkhoff
  theorem}~\cite{jebsen21,birkhoff23_relat_moder_physic,alexandrow23,eisland25}.



\subsubsection*{Calculating the field equations}

The Einstein equations in vacuum are equivalent to the vanishing
Ricci tensor, \(R_{\mu\nu} = 0\). Hence, we proceed to calculate the
Ricci tensor from the ansatz in Eq.~\eqref{eq:schw-ansatz}.

In \texttt{cadabra}, the metric is defined through a series of
substitutions, and the inverse metric is calculated---by
\texttt{SymPy} algorithms---from the same substitutions. For this end,
the algorithm \texttt{complete} is used. Notice below the use of the
\LaTeX{} notation for the exponential function.
\ContinueLineNumber
\begin{cadabra}
{A,B,f}::Depends(r).
ss := { g_{t t} = - \exp(A),
 g_{r r} = \exp(B),
 g_{\theta\theta} = r**2,
 g_{\phi\phi} = r**2 \sin(\theta)**2 }. 

complete(ss, $g^{\mu\nu}$);
\end{cadabra}
\begin{dmath*}[style={\small}]
  \left[g_{t t} = -\exp{A},~\discretionary{}{}{}
    g_{r r} = \exp{B},~\discretionary{}{}{}
    g_{\theta \theta} = {r}^{2},~\discretionary{}{}{}
    g_{\phi \phi} = {r}^{2} {\left(\sin{\theta}\right)}^{2},~\discretionary{}{}{}
    g^{t t} = -\exp\left(-A\right),~\discretionary{}{}{}
    g^{r r} = \exp\left(-B\right),~\discretionary{}{}{}
    g^{\phi \phi} = {\left({r}^{2} {\left(\sin{\theta}\right)}^{2}\right)}^{-1},~\discretionary{}{}{}
    g^{\theta \theta} = {r}^{-2}\right]
\end{dmath*}

With the metric (and its inverse) defined, one proceeds to calculate
the components of the Levi-Civita, Riemann and Ricci tensors, and the
scalar curvature. This is performed by the algorithm
\texttt{evaluate}, which accepts as second argument a series of
substitution rules to execute the evaluation.  For the Levi-Civita
connection the rules are the components of the metric,
i.e. \texttt{ss}.
\begin{cadabra}
evaluate(ch(), ss, rhsonly=True);
\end{cadabra}
\begin{dmath*}[style={\small}]
  \Gamma^{\mu}\,_{\nu \tau} = \square{}_{\nu}{}_{\tau}{}^{\mu}\left\{\begin{aligned}\square{}_{\phi}{}_{r}{}^{\phi} & = {r}^{-1}\\[-.5ex]
      \square{}_{\phi}{}_{\theta}{}^{\phi} & = {\left(\tan{\theta}\right)}^{-1}\\[-.5ex]
      \square{}_{\theta}{}_{r}{}^{\theta} & = {r}^{-1}\\[-.5ex]
      \square{}_{r}{}_{r}{}^{r} & = \frac{1}{2}\partial_{r}{B}\\[-.5ex]
      \square{}_{t}{}_{r}{}^{t} & = \frac{1}{2}\partial_{r}{A}\\[-.5ex]
      \square{}_{r}{}_{\phi}{}^{\phi} & = {r}^{-1}\\[-.5ex]
      \square{}_{\theta}{}_{\phi}{}^{\phi} & = {\left(\tan{\theta}\right)}^{-1}\\[-.5ex]
      \square{}_{r}{}_{\theta}{}^{\theta} & = {r}^{-1}\\[-.5ex]
      \square{}_{r}{}_{t}{}^{t} & = \frac{1}{2}\partial_{r}{A}\\[-.5ex]
      \square{}_{\phi}{}_{\phi}{}^{r} & = -r \exp\left(-B\right) {\left(\sin{\theta}\right)}^{2}\\[-.5ex]
      \square{}_{\phi}{}_{\phi}{}^{\theta} & =  - \frac{1}{2}\sin\left(2\theta\right)\\[-.5ex]
      \square{}_{\theta}{}_{\theta}{}^{r} & = -r \exp\left(-B\right)\\[-.5ex]
      \square{}_{t}{}_{t}{}^{r} & = \frac{1}{2}\exp\left(A-B\right) \partial_{r}{A}\\[-.5ex]
    \end{aligned}\right.
\end{dmath*}

Since the Riemannian curvature is expressed in terms of the
connection, one should first substitute the algebraic expression of
the Levi-Civita connection in terms of the metric (using the
\texttt{substitute} algorithm), and finally evaluate the curvature by
passing the substitution rules of the metric.
\begin{cadabra}
rm_comp = substitute(rm(), ch()) 
evaluate(rm_comp, ss, rhsonly=True);
\end{cadabra}
The output of the \texttt{evaluate} command is large, and useless for
our purpose in this section. Therefore, in the following some output
lines are avoided intentionally.

The Ricci tensor is defined from the Riemann tensor, one should
substitute its expression (notice that at this stage the Riemann
tensor is defined in terms of the metric, its derivative and the
metric inverse) before calculating with the algorithm \texttt{evaluate}.
\begin{cadabra}
rc_comp = substitute(rc(), rm) 
evaluate(rc_comp, ss, rhsonly=True);
\end{cadabra}
\begin{dmath*}
  R_{\sigma \nu} = \square{}_{\nu}{}_{\sigma}\left\{\begin{aligned}\square{}_{t}{}_{t} & = \left(r \left(\frac{1}{4}{\left(\partial_{r}{A}\right)}^{2} - \frac{1}{4}\partial_{r}{A} \partial_{r}{B}+\frac{1}{2}\partial_{r r}{A}\right)+\partial_{r}{A}\right) \exp\left(A-B\right) {r}^{-1}\\[-.5ex]
      \square{}_{\theta}{}_{\theta} & = \frac{1}{2}\left(-r \partial_{r}{A}+r \partial_{r}{B}+2\exp{B}-2\right) \exp\left(-B\right)\\[-.5ex]
      \square{}_{\phi}{}_{\phi} & = \frac{1}{2}\left(-r \partial_{r}{A}+r \partial_{r}{B}+2\exp{B}-2\right) \exp\left(-B\right) {\left(\sin{\theta}\right)}^{2}\\[-.5ex]
      \square{}_{r}{}_{r} & = \left(r \left( - \frac{1}{4}{\left(\partial_{r}{A}\right)}^{2}+\frac{1}{4}\partial_{r}{A} \partial_{r}{B} - \frac{1}{2}\partial_{r r}{A}\right)+\partial_{r}{B}\right) {r}^{-1}\\[-.5ex]
    \end{aligned}\right.
\end{dmath*}

\subsubsection*{Solving the field equations}

At the end of last section we manage to calculate the Ricci tensor
derived from the Schwarzschild metric anzats. Since the Einstein
equations in vacuum (without cosmological constant) are equivalent to
vanishing Ricci tensor, in this section we shall manipulate the
results above to solve the field equations.

The package \texttt{cdb.core.component}---imported in the header 
file---defines functions that allow you to access components of a 
\emph{geometrical object}. In particular, the function 
\texttt{get_component}, extracts a single component of an
expression (aka geometrical object).

Now, we can assign the calculated components to expressions, say
\(r00\) and \(r11\).
\begin{cadabra}
r00 = comp.get_component(rc_comp, $t,t$)[1];
r11 = comp.get_component(rc_comp, $r,r$)[1];
\end{cadabra}
\begin{dgroup*}
  \begin{dmath*}{}
    \left(r \left(\frac{1}{4}{\left(\partial_{r}{A}\right)}^{2} - \frac{1}{4}\partial_{r}{A} \partial_{r}{B}+\frac{1}{2}\partial_{r r}{A}\right)+\partial_{r}{A}\right) \exp\left(A-B\right) {r}^{-1}
  \end{dmath*}
  \begin{dmath*}
    \left(r \left( - \frac{1}{4}{\left(\partial_{r}{A}\right)}^{2}+\frac{1}{4}\partial_{r}{A} \partial_{r}{B} - \frac{1}{2}\partial_{r r}{A}\right)+\partial_{r}{B}\right) {r}^{-1}
  \end{dmath*}
\end{dgroup*}

The symbol \verb|[1]| at the end of the \texttt{get\_component}
function restricts our selection to the right-hand side of the
expression. 

Consider the  combination
\begin{equation*}
r \left( \exp(B - A) R_{t t} + R_{r r} \right) = 0.
\end{equation*}
\begin{cadabra}
expr1 := r \exp(B - A) @(r00) + r @(r11);
\end{cadabra}

There are several ways of simplifying the above result. Below we use
the function \texttt{map\_sympy} to expand the expression.
\begin{cadabra}
map_sympy(expr1, "expand");
\end{cadabra}
\begin{equation*}
\partial_{r}{A}+\partial_{r}{B}
\end{equation*}

Then the first expression, i.e. \texttt{expr1}, requires that
\(B(r) = - A(r) + C\). For the sake of simplicity, we shall set
\(C = 0\), nonetheless, in general grounds this choice represents a
scaling on the time coordinate.

Now consider the \(\theta \theta\)-component of the Ricci tensor,
after substituting \(B(r) = - A(r)\).
\begin{cadabra}
r22 = comp.get_component(rc, $\theta, \theta$)[1];
expr2 := @(r22):
substitute(expr2, $B -> - A$);
\end{cadabra}

The ordinary differential equation looks nicer if one transform even
further \({\exp(A) \to f}\) or equivalently \({A \to \ln(f)}\).
\begin{cadabra}
substitute(_ , $A -> \log(f)$)
map_sympy(_ , "expand");
\end{cadabra}
\begin{equation*}
-r \partial_{r}{f}-f+1
\end{equation*}

The \texttt{cdb.sympy.solvers} package provides wrappers to make
\texttt{SymPy} solvers available from Cadabra. It was imported in the
\texttt{header}, therefore, we use the algorithm \texttt{dsolve} to
solve the differential equation.
\begin{cadabra}
eq2 := @(expr2) = 0;
solA = solv.dsolve(eq2, $f$);
\end{cadabra}
\begin{equation*}
  \begin{split}
    & -r \partial_{r}{f}-f+1 = 0
    \\
    & f = C1 {r}^{-1}+1
  \end{split}
\end{equation*}

Therefore, the (static) spherically symmetric Ricci flat solution of Einstein field equations
requires the metric tensor to be
\begin{equation*}
\mathrm{d}s^2(g) = - \bigg( 1 + \frac{C_1}{r} \bigg)  \mathrm{d}t^2 + \frac{\mathrm{d}r^2}{ \Big( 1 + \frac{C_1}{r} \Big) }
+ r^2 \mathrm{d}\Omega_{(2)}^2.
\end{equation*}
The constant \(C_1\) is fixed to \(-2m\) by the Newtonian limit.

\subsection{Friedman--Robertson--Walker spacetime}
\label{sec:frw}

The Friedman--Robertson--Walker spacetime is a model of the Universe
with assumes the dubbed \emph{cosmological principle}, i.e. the
Universe at each time should look the same in all directions and also
to any fundamental observer. The cosmological principle causes that
the space is homogeneous (spacial translations are isometries) and
isotropic (spacial rotations are also isometries).

The metric compatible with the cosmological principle has a line
element given by (see for example Ref.~\cite{weinberg08_cosmol})
\begin{equation*}
  \mathrm{d}s^2(g) = - \mathrm{d}t^2 + a(t)^2 \frac{\mathrm{d}r^2}{1 - \kappa
    r^2} + a(t)^2 r^2 \mathrm{d}\Omega_{(2)}^2.
\end{equation*}

\subsubsection*{Calculating the field equations}

In the process of finding the field equations for the
Friedman--Robertson--Walker spacetimes, we repeat several steps
already mentioned in the Sec.~\ref{sec:schw}, hence we shall focus in
explaining the novelties.

We define three functions of time (scale factor, mass density and
pressure), and the rules for assigning the metric and the Kronecker
delta---used later to assign the components of the energy-momentum
tensor. Finally, we \texttt{complete} the inverse metric.
\StartLineAt{3}
\begin{cadabra}
{a,\rho,p}::Depends(t).

dl := { \delta_{t}^{t} = 1,
 \delta_{r}^{r} = 1,
 \delta_{\theta}^{\theta} = 1,
 \delta_{\phi}^{\phi} = 1 }. 

ss := { g_{t t} = - 1,
 g_{r r} = a**2 / (1 - k r**2),
 g_{\theta\theta} = a**2 r**2,
 g_{\phi\phi} = a**2 r**2 \sin(\theta)**2 }. 

complete(ss, $g^{\mu\nu}$);
\end{cadabra}
\begin{dmath*}[style={\small}]
  \left[g_{t t} = -1,~\discretionary{}{}{} g_{r r} = {a}^{2} {\left(1-k {r}^{2}\right)}^{-1},~\discretionary{}{}{} g_{\theta \theta} = {a}^{2} {r}^{2},~\discretionary{}{}{} g_{\phi \phi} = {a}^{2} {r}^{2} {\left(\sin{\theta}\right)}^{2},~\discretionary{}{}{} g^{t t} = -1,~\discretionary{}{}{} g^{r r} = \left(-k {r}^{2}+1\right) {a}^{-2},~\discretionary{}{}{} g^{\theta \theta} = {\left({r}^{2} {a}^{2}\right)}^{-1},~\discretionary{}{}{} g^{\phi \phi} = {\left({r}^{2} {a}^{2} {\left(\sin{\theta}\right)}^{2}\right)}^{-1}\right]
\end{dmath*}
Then, we calculate the Levi-Civita connection and curvature tensors
(most of the outputs returned by the code below are omitted).
\ContinueLineNumber
\begin{cadabra}
ch_comp = evaluate(ch(), ss, rhsonly=True);
rm_comp = substitute(rm(), ch_comp)
evaluate(rm_comp, ss, rhsonly=True);
rc_comp = substitute(rc(), rm_comp)
evaluate(rc_comp, ss, rhsonly=True);
rs_comp = substitute(rs(), rc_comp)
evaluate(rs_comp, ss, rhsonly=True);
ei_comp = substitute(ei(), rc_comp);
substitute(_, rs_comp);
evaluate(_, ss, rhsonly = True);
\end{cadabra}
\begin{dmath*}
G_{\mu \nu} = \square{}_{\nu}{}_{\mu}\left\{\begin{aligned}\square{}_{r}{}_{r} & = \left(k+2a \partial_{t t}{a}+{\left(\partial_{t}{a}\right)}^{2}\right) {\left(k {r}^{2}-1\right)}^{-1}\\[-.5ex]
\square{}_{\theta}{}_{\theta} & = -{r}^{2} \left(k+2a \partial_{t t}{a}+{\left(\partial_{t}{a}\right)}^{2}\right)\\[-.5ex]
\square{}_{\phi}{}_{\phi} & = -{r}^{2} \left(k+2a \partial_{t t}{a}+{\left(\partial_{t}{a}\right)}^{2}\right) {\left(\sin{\theta}\right)}^{2}\\[-.5ex]
\square{}_{t}{}_{t} & = 3\left(k+{\left(\partial_{t}{a}\right)}^{2}\right) {a}^{-2}\\[-.5ex]
\end{aligned}\right.
\end{dmath*}
It is useful to work the field equations with mixed indices, so we
proceed to define the \(\binom{1}{1}\)-type Einstein tensor.
\begin{cadabra}
Gud := ei_{\mu}^{\nu} = ei_{\mu \sigma} g^{\sigma \nu};
substitute(_, ei_comp)
evaluate(_, ss, rhsonly = True);
\end{cadabra}
\begin{dmath*}
G_{\mu}\,^{\nu} = \square{}_{\mu}{}^{\nu}\left\{\begin{aligned}\square{}_{r}{}^{r} & = -\left(k+2a \partial_{t t}{a}+{\left(\partial_{t}{a}\right)}^{2}\right) {a}^{-2}\\[-.5ex]
\square{}_{\theta}{}^{\theta} & = -\left(k+2a \partial_{t t}{a}+{\left(\partial_{t}{a}\right)}^{2}\right) {a}^{-2}\\[-.5ex]
\square{}_{\phi}{}^{\phi} & = -\left(k+2a \partial_{t t}{a}+{\left(\partial_{t}{a}\right)}^{2}\right) {a}^{-2}\\[-.5ex]
\square{}_{t}{}^{t} & = -\left(3k+3{\left(\partial_{t}{a}\right)}^{2}\right) {a}^{-2}\\[-.5ex]
\end{aligned}\right.
\end{dmath*}

In General Relativity a perfect fluid is modelled by a energy-momentum
tensor given by
\begin{equation}
  T^{\mu\nu} = (\rho + p) u^{\mu} u^{\nu} + p g^{\mu\nu},
\end{equation}
with \(u^\mu\) the \(4\)-velocity. Therefore, we define the substitution
rule that assign the components of the \(4\)-velocity in the comoving
frame, and then use this definition to evaluate the components of
the energy-momentum tensor.
\begin{cadabra}
u := [
  u^t = 1,
  u^r = 0,
  u^\theta = 0,
  u^\phi = 0
]:
T := T^{\mu\nu} = (\rho + p) u^{\mu} u^{\nu} + p g^{\mu\nu};
evaluate(T, join(u,ss), rhsonly=True);
\end{cadabra}
And the energy-momentum tensor with mixed indices.
\begin{cadabra}
Tud := T_{\mu}^{\nu} = g_{\mu \lambda} T^{\lambda \nu}:
substitute(Tud, T)
evaluate(Tud, join(u,ss), rhsonly=True);
\end{cadabra}
\begin{dmath*}
T_{\mu}\,^{\nu} = \square{}_{\mu}{}^{\nu}\left\{\begin{aligned}\square{}_{t}{}^{t} & = -\rho\\[-.5ex]
\square{}_{r}{}^{r} & = p\\[-.5ex]
\square{}_{\theta}{}^{\theta} & = p\\[-.5ex]
\square{}_{\phi}{}^{\phi} & = p\\[-.5ex]
\end{aligned}\right.
\end{dmath*}
Notice that in order to evaluate the components of the energy-momentum
tensor, we need the components of both the metric (\verb+ss+) and the
velocity (\verb+u+), and the way to call them together is through the
use of the \verb+join+ function, i.e. \verb+join(u,ss)+.\footnote{The
  \texttt{join} function was introduced in version \texttt{2.3.9.4}, and
  might be interchanged by a \emph{tilde concatenation}, e.g.
  \texttt{@(u) \~{} @(ss)} (See the discussion in the
  \href{https://cadabra.science/qa/2296/problem-with-changed-behaviour-of-for-lists}{cadabra
    Q\&A page}). In previous releases the substitution rules were
  called together with the plus sign, i.e. \texttt{u+ss}.}

Then, the Einstein field equations are
\begin{cadabra}
einseq := EQ_{\mu}^{\nu} = G_{\mu}^{\nu}  + \Lambda \delta_{\mu}^{\nu} - 8 \pi G T_{\mu}^{\nu};
substitute(einseq, Tud)
substitute(einseq, Gud)
evaluate(einseq, join(u,join(ss,dl)), rhsonly = True);
\end{cadabra}
\begin{dmath*}
EQ_{\mu}\,^{\nu} = \square{}_{\mu}{}^{\nu}\left\{\begin{aligned}\square{}_{r}{}^{r} & = -8\pi G p+\Lambda-k {a}^{-2}-2\partial_{t t}{a} {a}^{-1}-{\left(\partial_{t}{a}\right)}^{2} {a}^{-2}\\[-.5ex]
\square{}_{\theta}{}^{\theta} & = -8\pi G p+\Lambda-k {a}^{-2}-2\partial_{t t}{a} {a}^{-1}-{\left(\partial_{t}{a}\right)}^{2} {a}^{-2}\\[-.5ex]
\square{}_{\phi}{}^{\phi} & = -8\pi G p+\Lambda-k {a}^{-2}-2\partial_{t t}{a} {a}^{-1}-{\left(\partial_{t}{a}\right)}^{2} {a}^{-2}\\[-.5ex]
\square{}_{t}{}^{t} & = 8\pi G \rho+\Lambda-3k {a}^{-2}-3{\left(\partial_{t}{a}\right)}^{2} {a}^{-2}\\[-.5ex]
\end{aligned}\right.
\end{dmath*}

\subsubsection*{Solving the field equations}

Now, we can assign the calculated components to expressions, say
\(eq00\) and \(eq11\), which allows us to define the Friedman
equations, \(F1\) and \(F2\).
\ContinueLineNumber
\begin{cadabra}
eq00 = comp.get_component(einseq, $t,t$)[1];
distribute(_)
F1 := @(eq00) = 0;
eq11 = comp.get_component(einseq, $r,r$)[1];
distribute(_)
F2 := (@(eq00) - 3 @(eq11))/2 = 0;
\end{cadabra}
\begin{dgroup*}
  \begin{dmath*}{}8\pi G \rho+\Lambda-3k {a}^{-2}-3{\left(\partial_{t}{a}\right)}^{2} {a}^{-2} = 0\end{dmath*}
  \begin{dmath*}{}4\pi G \rho-\Lambda+12\pi G p+3\partial_{t t}{a} {a}^{-1} = 0\end{dmath*}
\end{dgroup*}

Finally, we have to import the \texttt{cdb.sympy.solvers} package to
solve the field equations.

\begin{itemize}
\item \textbf{Friedmann cosmologies in vacuum:}
We first consider Robertson--Walker cosmologies that are also vacuum solutions, i.e.
\begin{equation*}
  \rho = p = 0.
\end{equation*}
Then, 
\begin{cadabra}
vac := {\rho = 0, p = 0};
vac1 := @(F1):
substitute(vac1, vac);
vac2 := @(F2):
substitute(vac2, vac);
\end{cadabra}
\begin{dgroup*}[noalign]
  \begin{dmath*}
    \left[\rho = 0, p = 0 \right]
  \end{dmath*}
  \begin{dmath*}
    \Lambda - 3 k a^{-2} - 3 (\partial_t a)^2 a ^{-2} = 0
  \end{dmath*}
  \begin{dmath*}
    - \Lambda + 3 \partial_{tt}a \, a^{-1} = 0
  \end{dmath*}
\end{dgroup*}

The second vacuum equation can be obtained from the first one, and thus 
does not restrict the solutions. Consider the time-derivative of the 
first Fridmann equation,
\begin{equation*}
\Big(3 \dot{a}^2  + 3 k - \Lambda a^2 \Big)^\cdot
=
2 \Big( 3 \ddot{a} - \Lambda \Big) \dot{a} = 0.
\end{equation*}
The expression between the brackets is just the second Friedmann equation
in vacuum.

We then restrict ourselves to solve the first Friedmann equation. 
Additionally, we shall integrate the equation for different values \(k\).

\begin{itemize}
\item Solution for \(k = 0\):

\begin{cadabra}
tmp := @(vac1);
substitute(_, $k = 0$)
solvac1 = solv.dsolve(tmp, $a$);
\end{cadabra}
\begin{dmath*}
a = C1 {\exp\left(\sqrt{3} \sqrt{\Lambda} t\right)}^{ - \frac{1}{3}}
\end{dmath*}

The scale factor is given by the expression
\begin{equation*}
a(t) = C_1 \exp \bigg( \sqrt{\frac{\Lambda}{3}} t \bigg).
\end{equation*}
The above expression is well defined on the real number for \(\Lambda \ge 0\). 
For \(\Lambda = 0\) the scale factor is a constant (which can be normalised to one without
losing generality). 

\item Solution for \(k = -1\):

\begin{cadabra}
tmp := @(vac1);
substitute(_, $k = -1$)
solvac2 = solv.dsolve(tmp, $a$);
\end{cadabra}
\begin{dmath*}
a = -\sqrt{3} \sinh\left(\sqrt{\Lambda} \left(C1+\frac{1}{3}\sqrt{3} t\right)\right) {\left(\sqrt{\Lambda}\right)}^{-1}
\end{dmath*}

The scale factor is (the integration constant can be absorbed on the definition of \(t\),
and represents a shift in the origin of time)
\[a(t) =  \sqrt{\frac{3}{\Lambda}} \sinh \bigg( \sqrt{\frac{\Lambda}{3}} t \bigg). \]
This expression is well definied in the realm of real numbers for \(\Lambda \in \mathbb{R}\).
Define \(q = \sqrt{\frac{|\Lambda|}{3}}\), then 
\begin{equation*}
  a(t) =
  \begin{cases}
    \frac{\sinh (q t)}{q} &  \Lambda > 0 \\
    t & \Lambda = 0 \\
    \frac{\sin (q t)}{q} &  \Lambda < 0
  \end{cases}
\end{equation*}

\item Solution for \(k = 1\):

\begin{cadabra}
tmp := @(vac1);
substitute(_, $k = 1$)
solvac3 = solv.dsolve(tmp, $a$);
\end{cadabra}
The output looks horrible, but despite the strange output one notices
that a real solution, i.e. \(a^2 >0\), require \(\Lambda > 0\). In
that case,
\[a(t) = \frac{\cosh \big( q t \big)}{q},\]
with \(q\) defined above.

\item {General solutions with vanishing cosmological constant:}

\begin{cadabra}
tmp := @(vac1);
substitute(_, $\Lambda = 0$)
solvac0 = solv.dsolve(tmp, $a$);
\end{cadabra}
\begin{dmath*}
a = C1-t \sqrt{-k}
\end{dmath*}

The above, admits two types of solutions (mentioned already in the
previous sections), say \(a = C_1\) for \(k = 0\) and \(a = t\) for
\(k = -1\). Both solutions are just the Minkowski flat
spacetime. Although the first can be realised directly, the second
requires a re-parametrisation of the coordinates,
\begin{equation*}
  T = t \sqrt{ 1+ r^2} \text{ and } R = t r.
\end{equation*}
\end{itemize}

\item \textbf{Standard cosmological models:}
As in the previous case, it is possible to obtain a relation between
the first and second Friedmann equations. A short calculation shows
that
\begin{equation*}
  F2 - \frac{\dot{F1}}{2 \dot{a}} \Rightarrow \dot{\rho} + 3
  \frac{\dot{a}}{a} \Big( \rho + p \Big) = 0.
\end{equation*}
The latter is nothing but the conservation of the stress-energy
tensor.

The strategy to find solutions to the field equations is to assume an
equation of state---a relation between the energy density and the
pressure---, which we shall assume as linear, i.e. \(p = w \rho\) for
\(w \in \mathbb{R}\).

\begin{cadabra}
conserv := \partial_{t}{\rho} + 3 \partial_{t}{a} (\rho + p) / a = 0;
eq_state := p = w \rho;
substitute(conserv, eq_state)
map_sympy(conserv, "simplify");
\end{cadabra}
\begin{dmath*}
\left(3\left(w+1\right) \rho \partial_{t}{a}+a \partial_{t}{\rho}\right) {a}^{-1} = 0
\end{dmath*}

\begin{cadabra}
sol_rho = solv.dsolve(conserv, $\rho$);
\end{cadabra}
\begin{dmath*}
\rho = C1 \exp\left(-3w \log\left(a\right)\right) {a}^{-3}
\end{dmath*}
A by hand  simplification allows us to rewrite the last expression as
\begin{equation*}
  \rho = C_1 a ^{-3(1 + w)}.
\end{equation*}

We shall consider three cases:
\begin{itemize}
\item Cold (non-relativistic) matter:
also dubbed \emph{dust}, \(w = 0\) or \(p=0\).
\begin{cadabra}
tmp := @(F1);
substitute(tmp, eq_state)
substitute(tmp, sol_rho)
substitute(tmp, $w = 0$)
substitute(tmp, $\Lambda = 0$)
substitute(tmp, $k = 0$)
map_sympy(tmp, "simplify");
\end{cadabra}
\begin{dmath*}
\left(8\pi C1 G-3a {\left(\partial_{t}{a}\right)}^{2}\right) {a}^{-3} = 0
\end{dmath*}
\begin{cadabra}
solv.dsolve(tmp, $a$);
\end{cadabra}

In the above expression, the real solution is
\begin{equation*}
a = {\left(\sqrt{C1} \sqrt{G} \left( - \frac{3}{2}C1-\sqrt{6} \sqrt{\pi} t\right)\right)}^{\frac{2}{3}},
\end{equation*}
whose behaviour is
\begin{equation*}
a(t) \propto t^{2/3}.
\end{equation*}

\item Hot (relativistic) matter:
also known as \emph{radiation}, \(w = 1/3\) or \(p = \rho/3\).
\begin{cadabra}
tmp := @(F1);
substitute(tmp, eq_state)
substitute(tmp, sol_rho)
substitute(tmp, $w = 1/3$)
substitute(tmp, $\Lambda = 0$)
substitute(tmp, $k = 0$)
map_sympy(tmp, "simplify");
\end{cadabra}
\begin{dmath*}
\left(8\pi C1 G-3{a}^{2} {\left(\partial_{t}{a}\right)}^{2}\right) {a}^{-4} = 0
\end{dmath*}
\begin{cadabra}
solv.dsolve(tmp, $a$);
\end{cadabra}
\begin{dmath*}
\left[a =  - \frac{1}{3}\sqrt{6} \sqrt{-3C1-2\sqrt{6} \sqrt{\pi} t \sqrt{C1 G}},\; a = \frac{1}{3}\sqrt{6} \sqrt{-3C1-2\sqrt{6} \sqrt{\pi} t \sqrt{C1 G}}\right]
\end{dmath*}
In the above expression, the behaviour is
\begin{equation*}
a(t) \propto \sqrt{t}.
\end{equation*}

\item Vacuum energy:
also known as \emph{cosmological constant}, \(w = -1\) or \(p = - \rho\).
\begin{cadabra}
tmp := @(F1);
substitute(tmp, eq_state)
substitute(tmp, sol_rho)
substitute(tmp, $w = -1$)
substitute(tmp, $\Lambda = 0$)
substitute(tmp, $k = 0$)
map_sympy(tmp, "simplify");
\end{cadabra}
\begin{dmath*}
8\pi C1 G-3{\left(\partial_{t}{a}\right)}^{2} {a}^{-2} = 0
\end{dmath*}
\begin{cadabra}
solv.dsolve(tmp, $a$);
\end{cadabra}
\begin{dmath*}
a = C1 {\exp\left(\sqrt{6} \sqrt{\pi} t \sqrt{C1 G}\right)}^{ - \frac{2}{3}}
\end{dmath*}
In the above expression, the behaviour is
\begin{equation*}
a(t) \propto \exp{t}.
\end{equation*}
This is known as the de Sitter model.
\end{itemize}
\end{itemize}

\section{Discussion and conclusions}
\label{sec:concl}

Along the paper we have introduced (through examples) some of the
basic functionalities of \verb+Cadabra2+, focusing in applications to
the analysis of gravitational models. We believe that once the user
surpasses the initial barrier of the philosophy---in which all the
properties of the objects have to be declared---, the notation and
manipulation of these objects using the built in \texttt{algorithms}
is straightforward. Moreover, one can stack a sequence of these
algorithms into a custom function or even build algorithms from 
scratch, to ease the tasks during the process of solving problems.

In Sec.~\ref{sec:header} we mentioned the advantage of creating a
(\emph{per project}) header file, to facilitate the declaration of
variables. That file could---in principle--be shared between projects,
but we recommend to avoid such practice, since the personal notation
might change (depending for example of the collaborators of the
projects).

The tensor manipulation exemplified in Sec.~\ref{sec:manip-tens} shows
the simplicity behind the action of the algorithms on
expressions. Nonetheless, we should highlight the fact that in order
to achieve that simplicity, one has to relax the mathematical
formalities. Notice for example that in our example the Einstein sum
convention was dropped in favour of the simplicity of the substitution
rules.

More refined tensor manipulations were shown in
Sec.~\ref{sec:varprinc}, in which we used widely the substitution of
expressions, and the library \verb+cdb.core.manip+ to accomplish our
goal. It is important to make an allusion to the fact that within the
current version of the software, the \texttt{substitute} algorithm is
unable to distinguish the expression \verb+ch^{\rho}_{\mu \nu}+ from
\verb+\Gamma^{\rho}_{\mu \nu}+ (despite the fact that we specify the
\verb+\Gamma+ typography for the \verb+ch+ expression). This might
cause a undesired results.

We show the operational and computational capabilities of
\texttt{Cadabra2} in handling tensor components, and interact with the
\verb+Python+ library \verb+SymPy+. Through Sec.~\ref{sec:sols} we
managed to find and solve the Einstein field equations for the
Schwarzschild and Friedman--Robertson--Walker ansatz\"es. Although we
stick to \verb+SymPy+ as computational backend, if installed in your
system, you could use \verb+Mathematica+'s kernel as computational
backend.

In Sec.~\ref{sec:diff-forms} we showed a first application of the
ability of \verb+cadabra2+ to operate with differential forms. Even
though our example was very simple, it is possible to go further with
the current competence with the exterior calculus. However, there is
room to several improvements, as for example managing algebra valued
differential forms.

To conclude, we want to remark that given the extensible potential
of \verb+cadabra2+ through \verb+Python+, it is possible to use the
software to target research problems. Future work will focus on
applying the above concepts to the analysis of gravitational waves
\cite{castillo-felisola20_cadab_python_algor_gener_relat_cosmol_ii}.

\section*{Acknowledgements}

The authors wish to thank Kasper Peeters for developing and
maintaining \verb+cadabra2+ software, and Leo Brewin, who developed
the \texttt{cdblatex.sty} \LaTeX{} package that was used extensively
in the typesetting of this work~\cite{brewin19_hybrid-latex}.

MS would like to thank E.DeLazzari, S.Vidotto, A.Quaggio and
G.Casagrande for their shared passion and teachings. The work of OCF
is sponsored by the ``Centro Científico y Tecnológico de Valparaíso''
\mbox{(CCTVal)}, which is funded by the grant ANID PIA/APOYO AFB180002
(Chile). This research benefited from the grant \verb+PI_LI_19_02+
from the Universidad Técnica Federico Santa María.

\appendix

\section{Benchmarking in Cadabra}
\label{sec:bench}

Benchmarking is useful for a variety of purposes; locating
bottlenecks, optimisation, examining complexity, comparison again
other software packages: the list goes on. In perturbative
calculations many of these become relevant and so we deveoped a timing
module which is distributed with this paper.

The basic principle is simple: \texttt{Timer} objects measure the time
taken between reaching various \emph{checkpoints}, and accumulate
these into named bins. For example
\StartLineAt{1}
\begin{cadabra}
from libraries.timing import Timer

timer = Timer()
timer.start()
foo()
timer.checkpoint("foo")
boring()
timer.checkpoint()
bar()
timer.checkpoint("bar")
foo()
timer.checkpoint("foo")
\end{cadabra}
In this snippet we have three functions; \texttt{foo} and \texttt{bar}
which we want to measure, and \texttt{boring} which we need to run but
we aren't very interested in. After calling \texttt{start} on our
timer, we cycle through some \texttt{foo} routines then call
\texttt{checkpoint} with the ``foo'' label. We then do our
\texttt{boring} routine, and to avoid this cluttering up our final
timing results once we're done with this we call \texttt{checkpoint}
with an empty label. We then all and checkpoint \texttt{bar}, followed
by a second cycle of \texttt{foo}s which accumulates to the time the
first call to \texttt{checkpoint("foo")} registered.

We can have a look at the amount of time spent doing each routine by
calling \texttt{plot\_bar\_chart} which uses \texttt{matplotlib} to
produce a plot of the results:
\ContinueLineNumber
\begin{cadabra}
timer.plot_bar_chart();
\end{cadabra}

\begin{figure}[ht]
  \centering
  \includegraphics[width=.7\textwidth]{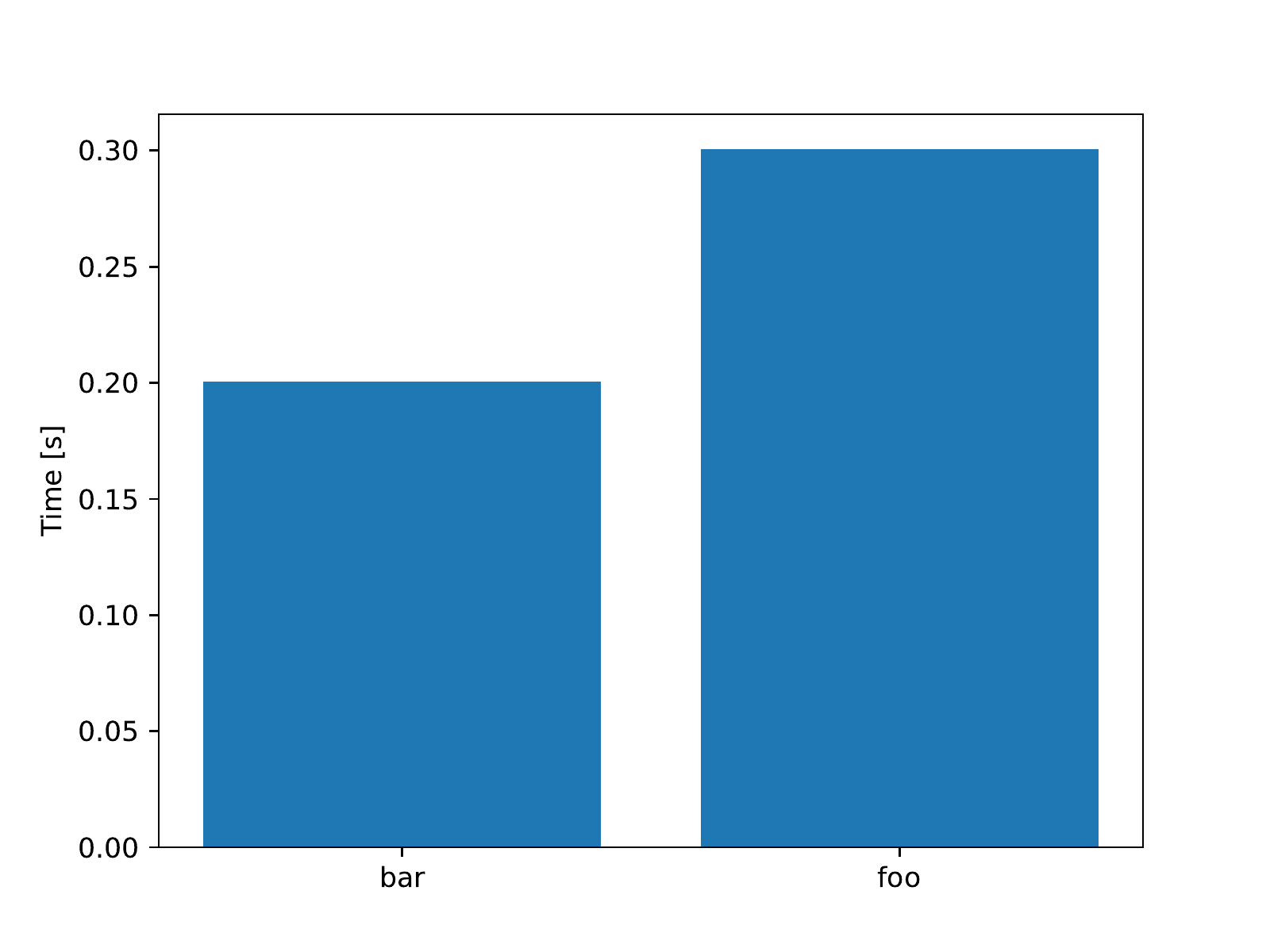}
\end{figure}

We can also see how long the entire routine took by calling
\texttt{total\_elapsed}. If we want to include the time spent in our
\texttt{boring} function, then passing \texttt{include\_idle=True}
adds this to the total:
\begin{cadabra}
print("timer.total_elapsed(False):", timer.total_elapsed(False))
print("timer.total_elapsed(True):", timer.total_elapsed(True))
\end{cadabra}
\begin{verbatim}
    timer.total_elapsed(False): 0.5009605884552002
    timer.total_elapsed(True): 1.5020785331726074
\end{verbatim}

To demonstrate an example where this can help us optimise a
calculation in Cadabra, consider the following code snippet:
\begin{cadabra}
coordinates := t, x, y, z.
@(coordinates)::Coordinate.
\partial{#}::PartialDerivative.
\Phi{#}::Depends(\partial{#}).

# Build expression
ex := 0.
for c1 in coordinates.top().children():
  for c2 in coordinates.top().children():
    ex += $\partial_{@(c1)}{ \Phi \partial_{@(c2)}{ \Phi } }$

subrule := \Phi -> \lambda_1 T(t) R_1(x,y,z) + \lambda_2 T(t) R_2(x,y,z).
\end{cadabra}

We have constructed an equation consisting of a sum over partial
derivatives, and we now wish to substitute in an expression for $\Phi$
(in this case we will use the illustrative nonsense
$\Phi \rightarrow \lambda_1 T(t) R_1(x,y,z) + \lambda_2 T(t)
R_2(x,y,z)$) ) and then expand out the result. This will involve using
the three algorithms \verb|distribute|, \verb|product_rule| and
\verb|unwrap|.

We perform this below and plot a graph of the time spent in each
algorithm. Note that we call \verb|checkpoint()| at the start of each
loop so that the time \verb|converge| spends calculating the loop
predicate is not included in the timing of the \verb|distribute|
algorithm.

\begin{cadabra}
t1 = Timer("dist-pr-unwrap")

test = substitute($@(ex)$, subrule)
t1.start()
converge(test):
	t1.checkpoint()
	distribute(test)
	t1.checkpoint("distribute")
	product_rule(test)
	t1.checkpoint("product_rule")
	unwrap(test)
	t1.checkpoint("unwrap")

t1.plot_bar_chart();
\end{cadabra}

\begin{figure}[ht]
  \centering
  \includegraphics[width=.7\textwidth]{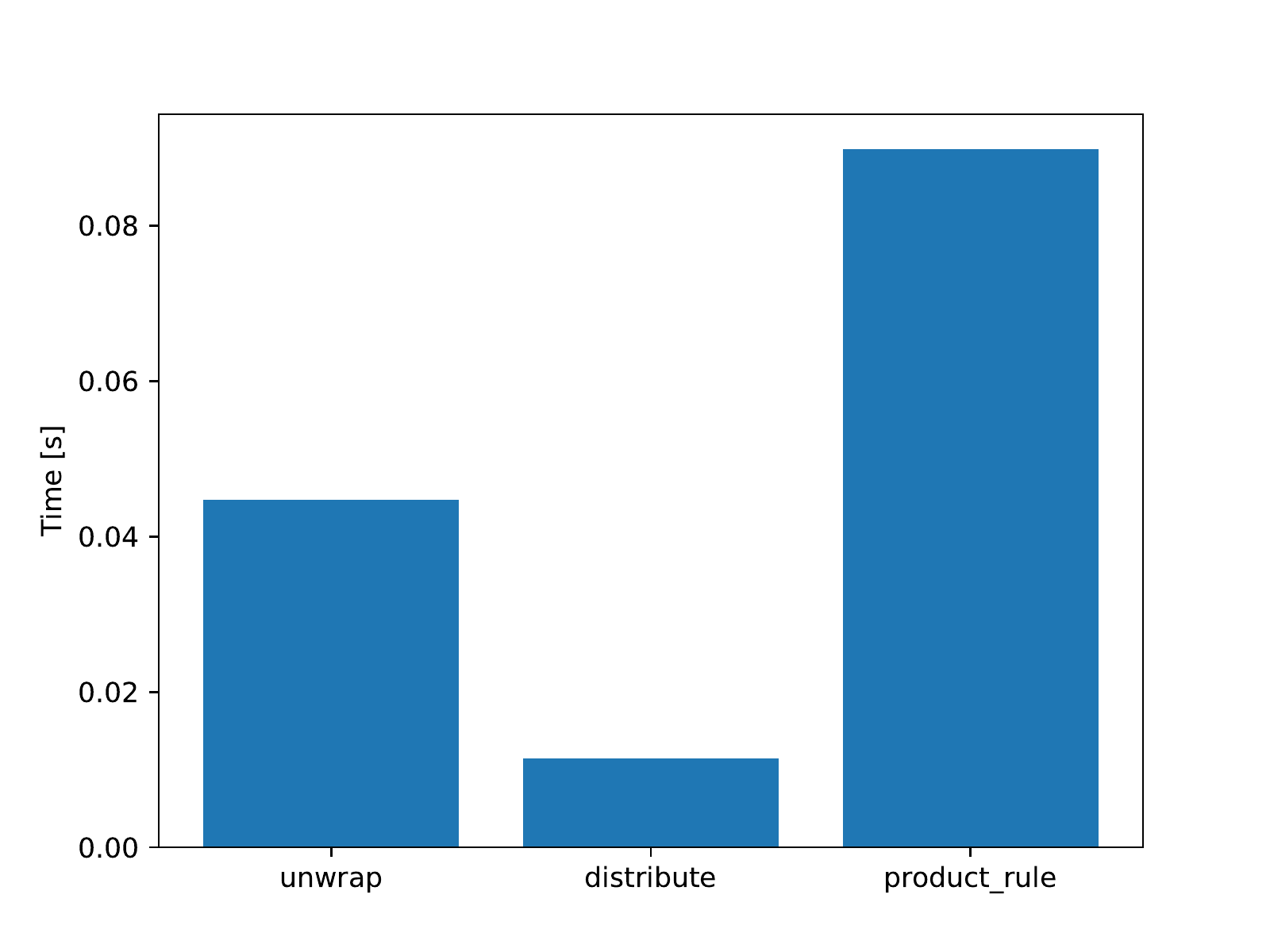}
\end{figure}

Clearly there is a bottleneck in the \verb|product_rule| algorithm,
and if we wished to repeat this calculation many times with different
substitutions we may want to try and find a more efficient approach to
the problem. We try out a couple of different orders to see if this
has an impact:
\begin{cadabra}
t2 = Timer("pr-dist-unwrap")

test = substitute($@(ex)$, subrule)
t2.start()
converge(test):
  t2.checkpoint()
  product_rule(test)
  t2.checkpoint("product_rule")
  distribute(test)
  t2.checkpoint("distribute")
  unwrap(test)
  t2.checkpoint("unwrap")

t3 = Timer("dist-unwrap-pr")

test = substitute($@(ex)$, subrule)
t3.start()
converge(test):
  t3.checkpoint()
  product_rule(test)
  t3.checkpoint("product_rule")
  distribute(test)
  t3.checkpoint("distribute")
  unwrap(test)
  t3.checkpoint("unwrap")        
\end{cadabra}

In order to be able to compare the time spent in each algorithm side
by side, we can call the \verb|plot_bar_chart| method with other
timers as parameters to do plot a comparison:
\begin{cadabra}
t1.plot_bar_chart(t2, t3, legend=True);
\end{cadabra}
\begin{figure}[ht]
  \centering
  \includegraphics[width=.7\textwidth]{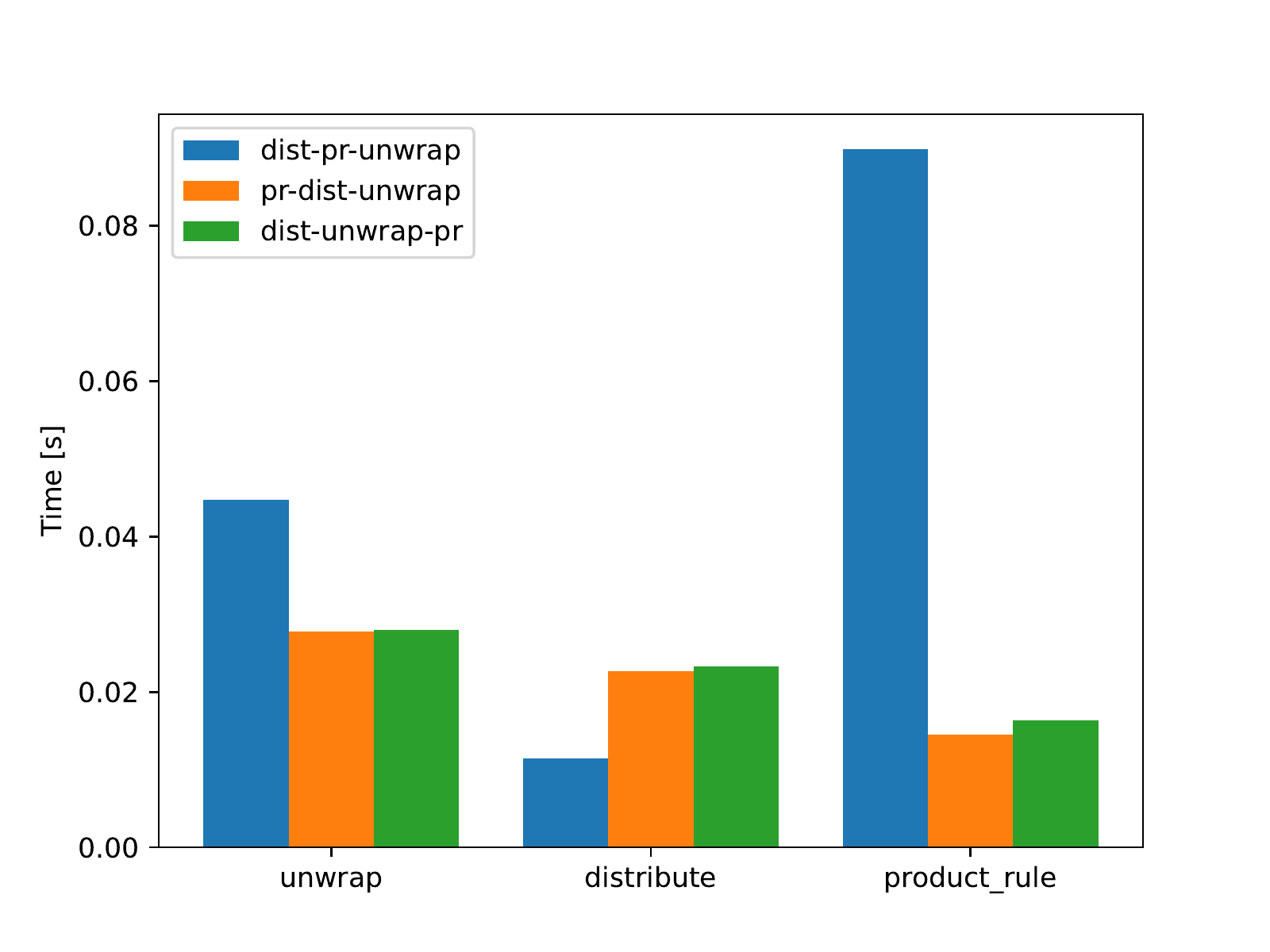}
\end{figure}
Clearly we can see that the orange and green bars representing the
second and third order are more efficient ways of performing the
calculation.


\section{Fourth-order curvature Lagrangian of the Lanczos--Lovelock series}
\label{sec:LL4}

As a complementary material, we present the term \(\Lag^{(D,4)}\) of
the Lanczos--Lovelock series. It is shown in a separated appendix
because obtaining it is not relevant for the purpose of the article,
however it could be useful for the reader interested in this research
topic. The code below follows from the developed in Sec.~\ref{sec:LL}.
\StartLineAt{41}
\begin{cadabra}
LL4 := 8*7*6*5*4*3*2/2/2/2/2 R^{a1 a2 b1 b2} R^{a3 a4 b3 b4} R^{a5 a6 b5 b6} R^{a7 a8 b7 b8} \delta^{a1 b1 a2 b2 a3 b3 a4 b4 a5 b5 a6 b6 a7 b7 a8 b8};
LLmanip(LL4);
\end{cadabra}
\begin{dgroup*}
  \begin{dmath*}
    {}2520R^{{a_{1}} {a_{2}} {b_{1}} {b_{2}}} R^{{a_{3}} {a_{4}} {b_{3}} {b_{4}}} R^{{a_{5}} {a_{6}} {b_{5}} {b_{6}}} R^{{a_{7}} {a_{8}} {b_{7}} {b_{8}}} \delta^{{a_{1}} {b_{1}} {a_{2}} {b_{2}} {a_{3}} {b_{3}} {a_{4}} {b_{4}} {a_{5}} {b_{5}} {a_{6}} {b_{6}} {a_{7}} {b_{7}} {a_{8}} {b_{8}}}
  \end{dmath*}
  \begin{dmath}
    {}{R}^{4}-24{R}^{2} R^{{a_{1}} {a_{2}}} R^{{a_{1}} {a_{2}}}+6{R}^{2} R^{{a_{1}} {a_{2}} {a_{3}} {a_{4}}} R^{{a_{1}} {a_{2}} {a_{3}} {a_{4}}}+64R R^{{a_{1}} {a_{2}}} R^{{a_{1}} {a_{3}}} R^{{a_{2}} {a_{3}}}+96R R^{{a_{1}} {a_{2}}} R^{{a_{3}} {a_{4}}} R^{{a_{1}} {a_{3}} {a_{2}} {a_{4}}}-96R R^{{a_{1}} {a_{2}}} R^{{a_{1}} {a_{3}} {a_{4}} {a_{5}}} R^{{a_{2}} {a_{3}} {a_{4}} {a_{5}}}+8R R^{{a_{1}} {a_{2}} {a_{3}} {a_{4}}} R^{{a_{1}} {a_{2}} {a_{5}} {a_{6}}} R^{{a_{3}} {a_{4}} {a_{5}} {a_{6}}}-32R R^{{a_{1}} {a_{2}} {a_{3}} {a_{4}}} R^{{a_{1}} {a_{5}} {a_{3}} {a_{6}}} R^{{a_{2}} {a_{6}} {a_{4}} {a_{5}}}+48R^{{a_{1}} {a_{2}}} R^{{a_{1}} {a_{2}}} R^{{a_{3}} {a_{4}}} R^{{a_{3}} {a_{4}}}-24R^{{a_{1}} {a_{2}}} R^{{a_{1}} {a_{2}}} R^{{a_{3}} {a_{4}} {a_{5}} {a_{6}}} R^{{a_{3}} {a_{4}} {a_{5}} {a_{6}}}-96R^{{a_{1}} {a_{2}}} R^{{a_{1}} {a_{3}}} R^{{a_{2}} {a_{4}}} R^{{a_{3}} {a_{4}}}-384R^{{a_{1}} {a_{2}}} R^{{a_{1}} {a_{3}}} R^{{a_{4}} {a_{5}}} R^{{a_{2}} {a_{4}} {a_{3}} {a_{5}}}+192R^{{a_{1}} {a_{2}}} R^{{a_{1}} {a_{3}}} R^{{a_{2}} {a_{4}} {a_{5}} {a_{6}}} R^{{a_{3}} {a_{4}} {a_{5}} {a_{6}}}+96R^{{a_{1}} {a_{2}}} R^{{a_{3}} {a_{4}}} R^{{a_{1}} {a_{3}} {a_{5}} {a_{6}}} R^{{a_{2}} {a_{4}} {a_{5}} {a_{6}}}-192R^{{a_{1}} {a_{2}}} R^{{a_{3}} {a_{4}}} R^{{a_{1}} {a_{5}} {a_{2}} {a_{6}}} R^{{a_{3}} {a_{5}} {a_{4}} {a_{6}}}+192R^{{a_{1}} {a_{2}}} R^{{a_{3}} {a_{4}}} R^{{a_{1}} {a_{5}} {a_{3}} {a_{6}}} R^{{a_{2}} {a_{5}} {a_{4}} {a_{6}}}+192R^{{a_{1}} {a_{2}}} R^{{a_{1}} {a_{3}} {a_{2}} {a_{4}}} R^{{a_{3}} {a_{5}} {a_{6}} {a_{7}}} R^{{a_{4}} {a_{5}} {a_{6}} {a_{7}}}-96R^{{a_{1}} {a_{2}}} R^{{a_{1}} {a_{3}} {a_{4}} {a_{5}}} R^{{a_{2}} {a_{3}} {a_{6}} {a_{7}}} R^{{a_{4}} {a_{5}} {a_{6}} {a_{7}}}+384R^{{a_{1}} {a_{2}}} R^{{a_{1}} {a_{3}} {a_{4}} {a_{5}}} R^{{a_{2}} {a_{6}} {a_{4}} {a_{7}}} R^{{a_{3}} {a_{7}} {a_{5}} {a_{6}}}+3R^{{a_{1}} {a_{2}} {a_{3}} {a_{4}}} R^{{a_{1}} {a_{2}} {a_{3}} {a_{4}}} R^{{a_{5}} {a_{6}} {a_{7}} {a_{8}}} R^{{a_{5}} {a_{6}} {a_{7}} {a_{8}}}-48R^{{a_{1}} {a_{2}} {a_{3}} {a_{4}}} R^{{a_{1}} {a_{2}} {a_{3}} {a_{5}}} R^{{a_{4}} {a_{6}} {a_{7}} {a_{8}}} R^{{a_{5}} {a_{6}} {a_{7}} {a_{8}}}+6R^{{a_{1}} {a_{2}} {a_{3}} {a_{4}}} R^{{a_{1}} {a_{2}} {a_{5}} {a_{6}}} R^{{a_{3}} {a_{4}} {a_{7}} {a_{8}}} R^{{a_{5}} {a_{6}} {a_{7}} {a_{8}}}-96R^{{a_{1}} {a_{2}} {a_{3}} {a_{4}}} R^{{a_{1}} {a_{2}} {a_{5}} {a_{6}}} R^{{a_{3}} {a_{7}} {a_{5}} {a_{8}}} R^{{a_{4}} {a_{8}} {a_{6}} {a_{7}}}+48R^{{a_{1}} {a_{2}} {a_{3}} {a_{4}}} R^{{a_{1}} {a_{5}} {a_{3}} {a_{6}}} R^{{a_{2}} {a_{7}} {a_{4}} {a_{8}}} R^{{a_{5}} {a_{7}} {a_{6}} {a_{8}}}-96R^{{a_{1}} {a_{2}} {a_{3}} {a_{4}}} R^{{a_{1}} {a_{5}} {a_{3}} {a_{6}}} R^{{a_{2}} {a_{7}} {a_{5}} {a_{8}}} R^{{a_{4}} {a_{7}} {a_{6}} {a_{8}}}
  \end{dmath}
\end{dgroup*}
while the field equations are given by
\begin{cadabra}
feqLL4 := - 9*8*7*6*5*4*3*2/2/2/2/2/2 R^{a1 a2 b1 b2} R^{a3 a4 b3 b4} R^{a5 a6 b5 b6} R^{a7 a8 b7 b8} \delta^{m n a1 b1 a2 b2 a3 b3 a4 b4 a5 b5 a6 b6 a7 b7 a8 b8};
LLmanip(feqLL4)
factor_out(_, $\delta^{m n}$)
substitute(_, $@(LL4) -> L^{(D,4)}$);
\end{cadabra}
\begin{dgroup*}
  \begin{dmath*}
    {}-11340R^{{a_{1}} {a_{2}} {b_{1}} {b_{2}}} R^{{a_{3}} {a_{4}} {b_{3}} {b_{4}}} R^{{a_{5}} {a_{6}} {b_{5}} {b_{6}}} R^{{a_{7}} {a_{8}} {b_{7}} {b_{8}}} \delta^{m n {a_{1}} {b_{1}} {a_{2}} {b_{2}} {a_{3}} {b_{3}} {a_{4}} {b_{4}} {a_{5}} {b_{5}} {a_{6}} {b_{6}} {a_{7}} {b_{7}} {a_{8}} {b_{8}}}
  \end{dmath*}
\end{dgroup*}
\begin{dmath}[style={\small}]
  G^{(4)}{}^{m n} =4{R}^{3} R^{m n}-24{R}^{2} R^{{a_{1}} {a_{2}}} R^{m {a_{1}} n {a_{2}}}-24{R}^{2} R^{m {a_{1}}} R^{n {a_{1}}}+12{R}^{2} R^{m {a_{1}} {a_{2}} {a_{3}}} R^{n {a_{1}} {a_{2}} {a_{3}}}+96R R^{n {a_{1}}} R^{{a_{2}} {a_{3}}} R^{m {a_{2}} {a_{1}} {a_{3}}}+96R R^{{a_{1}} {a_{2}}} R^{{a_{1}} {a_{3}}} R^{m {a_{2}} n {a_{3}}}-48R R^{{a_{1}} {a_{2}}} R^{m {a_{1}} {a_{3}} {a_{4}}} R^{n {a_{2}} {a_{3}} {a_{4}}}-48R R^{m n} R^{{a_{1}} {a_{2}}} R^{{a_{1}} {a_{2}}}+96R R^{m {a_{1}}} R^{n {a_{2}}} R^{{a_{1}} {a_{2}}}+96R R^{m {a_{1}}} R^{{a_{2}} {a_{3}}} R^{n {a_{2}} {a_{1}} {a_{3}}}+96R R^{{a_{1}} {a_{2}}} R^{m {a_{3}} n {a_{4}}} R^{{a_{1}} {a_{3}} {a_{2}} {a_{4}}}-96R R^{{a_{1}} {a_{2}}} R^{m {a_{3}} {a_{1}} {a_{4}}} R^{n {a_{3}} {a_{2}} {a_{4}}}-48R R^{m {a_{1}}} R^{n {a_{2}} {a_{3}} {a_{4}}} R^{{a_{1}} {a_{2}} {a_{3}} {a_{4}}}+12R R^{m n} R^{{a_{1}} {a_{2}} {a_{3}} {a_{4}}} R^{{a_{1}} {a_{2}} {a_{3}} {a_{4}}}-48R R^{n {a_{1}}} R^{m {a_{2}} {a_{3}} {a_{4}}} R^{{a_{1}} {a_{2}} {a_{3}} {a_{4}}}-48R R^{m {a_{1}} n {a_{2}}} R^{{a_{1}} {a_{3}} {a_{4}} {a_{5}}} R^{{a_{2}} {a_{3}} {a_{4}} {a_{5}}}-96R R^{m {a_{1}} {a_{2}} {a_{3}}} R^{n {a_{4}} {a_{2}} {a_{5}}} R^{{a_{1}} {a_{5}} {a_{3}} {a_{4}}}+24R R^{m {a_{1}} {a_{2}} {a_{3}}} R^{n {a_{1}} {a_{4}} {a_{5}}} R^{{a_{2}} {a_{3}} {a_{4}} {a_{5}}}-192R^{n {a_{1}}} R^{{a_{1}} {a_{2}}} R^{{a_{3}} {a_{4}}} R^{m {a_{3}} {a_{2}} {a_{4}}}+96R^{{a_{1}} {a_{2}}} R^{{a_{1}} {a_{2}}} R^{{a_{3}} {a_{4}}} R^{m {a_{3}} n {a_{4}}}-96R^{{a_{1}} {a_{2}}} R^{m {a_{1}} {a_{2}} {a_{3}}} R^{n {a_{4}} {a_{5}} {a_{6}}} R^{{a_{3}} {a_{4}} {a_{5}} {a_{6}}}-24R^{{a_{1}} {a_{2}}} R^{m {a_{1}} n {a_{2}}} R^{{a_{3}} {a_{4}} {a_{5}} {a_{6}}} R^{{a_{3}} {a_{4}} {a_{5}} {a_{6}}}-192R^{{a_{1}} {a_{2}}} R^{{a_{3}} {a_{4}}} R^{m {a_{1}} {a_{2}} {a_{5}}} R^{n {a_{3}} {a_{4}} {a_{5}}}-192R^{{a_{1}} {a_{2}}} R^{{a_{1}} {a_{3}}} R^{{a_{2}} {a_{4}}} R^{m {a_{3}} n {a_{4}}}+192R^{{a_{1}} {a_{2}}} R^{{a_{3}} {a_{4}}} R^{m {a_{1}} {a_{3}} {a_{5}}} R^{n {a_{2}} {a_{4}} {a_{5}}}+192R^{{a_{1}} {a_{2}}} R^{{a_{3}} {a_{4}}} R^{m {a_{1}} n {a_{5}}} R^{{a_{2}} {a_{3}} {a_{4}} {a_{5}}}-48R^{{a_{1}} {a_{2}}} R^{m {a_{1}} {a_{3}} {a_{4}}} R^{n {a_{2}} {a_{5}} {a_{6}}} R^{{a_{3}} {a_{4}} {a_{5}} {a_{6}}}+96R^{{a_{1}} {a_{2}}} R^{m {a_{1}} n {a_{3}}} R^{{a_{2}} {a_{4}} {a_{5}} {a_{6}}} R^{{a_{3}} {a_{4}} {a_{5}} {a_{6}}}-192R^{n {a_{1}}} R^{{a_{2}} {a_{3}}} R^{{a_{2}} {a_{4}}} R^{m {a_{3}} {a_{1}} {a_{4}}}+96R^{{a_{1}} {a_{2}}} R^{{a_{1}} {a_{3}}} R^{m {a_{2}} {a_{4}} {a_{5}}} R^{n {a_{3}} {a_{4}} {a_{5}}}+96R^{n {a_{1}}} R^{{a_{2}} {a_{3}}} R^{m {a_{2}} {a_{4}} {a_{5}}} R^{{a_{1}} {a_{3}} {a_{4}} {a_{5}}}+192R^{{a_{1}} {a_{2}}} R^{m {a_{1}} {a_{3}} {a_{4}}} R^{n {a_{5}} {a_{3}} {a_{6}}} R^{{a_{2}} {a_{6}} {a_{4}} {a_{5}}}+96R^{m {a_{1}}} R^{n {a_{1}}} R^{{a_{2}} {a_{3}}} R^{{a_{2}} {a_{3}}}-48R^{{a_{1}} {a_{2}}} R^{{a_{1}} {a_{2}}} R^{m {a_{3}} {a_{4}} {a_{5}}} R^{n {a_{3}} {a_{4}} {a_{5}}}+64R^{m n} R^{{a_{1}} {a_{2}}} R^{{a_{1}} {a_{3}}} R^{{a_{2}} {a_{3}}}-192R^{m {a_{1}}} R^{n {a_{2}}} R^{{a_{1}} {a_{3}}} R^{{a_{2}} {a_{3}}}-192R^{m {a_{1}}} R^{{a_{2}} {a_{3}}} R^{{a_{2}} {a_{4}}} R^{n {a_{3}} {a_{1}} {a_{4}}}-192R^{{a_{1}} {a_{2}}} R^{{a_{1}} {a_{3}}} R^{m {a_{4}} n {a_{5}}} R^{{a_{2}} {a_{4}} {a_{3}} {a_{5}}}+192R^{{a_{1}} {a_{2}}} R^{{a_{1}} {a_{3}}} R^{m {a_{4}} {a_{2}} {a_{5}}} R^{n {a_{4}} {a_{3}} {a_{5}}}-192R^{m {a_{1}}} R^{{a_{1}} {a_{2}}} R^{{a_{3}} {a_{4}}} R^{n {a_{3}} {a_{2}} {a_{4}}}+96R^{m n} R^{{a_{1}} {a_{2}}} R^{{a_{3}} {a_{4}}} R^{{a_{1}} {a_{3}} {a_{2}} {a_{4}}}+192R^{{a_{1}} {a_{2}}} R^{{a_{3}} {a_{4}}} R^{m {a_{5}} n {a_{1}}} R^{{a_{2}} {a_{3}} {a_{4}} {a_{5}}}+96R^{{a_{1}} {a_{2}}} R^{{a_{3}} {a_{4}}} R^{m {a_{5}} {a_{1}} {a_{3}}} R^{n {a_{5}} {a_{2}} {a_{4}}}+96R^{m {a_{1}}} R^{{a_{1}} {a_{2}}} R^{n {a_{3}} {a_{4}} {a_{5}}} R^{{a_{2}} {a_{3}} {a_{4}} {a_{5}}}-192R^{m {a_{1}}} R^{n {a_{2}}} R^{{a_{3}} {a_{4}}} R^{{a_{1}} {a_{3}} {a_{2}} {a_{4}}}-192R^{m {a_{1}}} R^{{a_{2}} {a_{3}}} R^{n {a_{4}} {a_{1}} {a_{5}}} R^{{a_{2}} {a_{4}} {a_{3}} {a_{5}}}+96R^{m {a_{1}}} R^{{a_{2}} {a_{3}}} R^{n {a_{2}} {a_{4}} {a_{5}}} R^{{a_{1}} {a_{3}} {a_{4}} {a_{5}}}+192R^{m {a_{1}}} R^{{a_{2}} {a_{3}}} R^{n {a_{4}} {a_{2}} {a_{5}}} R^{{a_{1}} {a_{4}} {a_{3}} {a_{5}}}-96R^{m n} R^{{a_{1}} {a_{2}}} R^{{a_{1}} {a_{3}} {a_{4}} {a_{5}}} R^{{a_{2}} {a_{3}} {a_{4}} {a_{5}}}+96R^{n {a_{1}}} R^{{a_{1}} {a_{2}}} R^{m {a_{3}} {a_{4}} {a_{5}}} R^{{a_{2}} {a_{3}} {a_{4}} {a_{5}}}-192R^{n {a_{1}}} R^{{a_{2}} {a_{3}}} R^{m {a_{4}} {a_{1}} {a_{5}}} R^{{a_{2}} {a_{4}} {a_{3}} {a_{5}}}+192R^{n {a_{1}}} R^{{a_{2}} {a_{3}}} R^{m {a_{4}} {a_{2}} {a_{5}}} R^{{a_{1}} {a_{4}} {a_{3}} {a_{5}}}+96R^{{a_{1}} {a_{2}}} R^{m {a_{3}} {a_{4}} {a_{5}}} R^{n {a_{6}} {a_{4}} {a_{5}}} R^{{a_{1}} {a_{3}} {a_{2}} {a_{6}}}+96R^{{a_{1}} {a_{2}}} R^{m {a_{3}} n {a_{4}}} R^{{a_{1}} {a_{3}} {a_{5}} {a_{6}}} R^{{a_{2}} {a_{4}} {a_{5}} {a_{6}}}-192R^{{a_{1}} {a_{2}}} R^{m {a_{3}} n {a_{4}}} R^{{a_{1}} {a_{5}} {a_{2}} {a_{6}}} R^{{a_{3}} {a_{5}} {a_{4}} {a_{6}}}+192R^{{a_{1}} {a_{2}}} R^{m {a_{3}} {a_{4}} {a_{5}}} R^{n {a_{3}} {a_{4}} {a_{6}}} R^{{a_{1}} {a_{5}} {a_{2}} {a_{6}}}+192R^{{a_{1}} {a_{2}}} R^{m {a_{3}} n {a_{4}}} R^{{a_{1}} {a_{5}} {a_{3}} {a_{6}}} R^{{a_{2}} {a_{5}} {a_{4}} {a_{6}}}+192R^{{a_{1}} {a_{2}}} R^{m {a_{3}} {a_{4}} {a_{5}}} R^{n {a_{6}} {a_{1}} {a_{4}}} R^{{a_{2}} {a_{3}} {a_{5}} {a_{6}}}+96R^{{a_{1}} {a_{2}}} R^{m {a_{3}} n {a_{1}}} R^{{a_{2}} {a_{4}} {a_{5}} {a_{6}}} R^{{a_{3}} {a_{4}} {a_{5}} {a_{6}}}-96R^{{a_{1}} {a_{2}}} R^{m {a_{3}} {a_{1}} {a_{4}}} R^{n {a_{3}} {a_{5}} {a_{6}}} R^{{a_{2}} {a_{4}} {a_{5}} {a_{6}}}-192R^{{a_{1}} {a_{2}}} R^{m {a_{3}} {a_{1}} {a_{4}}} R^{n {a_{5}} {a_{4}} {a_{6}}} R^{{a_{2}} {a_{5}} {a_{3}} {a_{6}}}-192R^{{a_{1}} {a_{2}}} R^{m {a_{3}} {a_{4}} {a_{5}}} R^{n {a_{1}} {a_{4}} {a_{6}}} R^{{a_{2}} {a_{5}} {a_{3}} {a_{6}}}-96R^{{a_{1}} {a_{2}}} R^{m {a_{3}} {a_{4}} {a_{5}}} R^{n {a_{3}} {a_{1}} {a_{6}}} R^{{a_{2}} {a_{6}} {a_{4}} {a_{5}}}+192R^{{a_{1}} {a_{2}}} R^{m {a_{3}} {a_{1}} {a_{4}}} R^{n {a_{5}} {a_{2}} {a_{6}}} R^{{a_{3}} {a_{6}} {a_{4}} {a_{5}}}+96R^{{a_{1}} {a_{2}}} R^{m {a_{3}} {a_{4}} {a_{5}}} R^{n {a_{1}} {a_{2}} {a_{6}}} R^{{a_{3}} {a_{6}} {a_{4}} {a_{5}}}-24R^{m {a_{1}}} R^{n {a_{1}}} R^{{a_{2}} {a_{3}} {a_{4}} {a_{5}}} R^{{a_{2}} {a_{3}} {a_{4}} {a_{5}}}+96R^{m {a_{1}}} R^{n {a_{2}}} R^{{a_{1}} {a_{3}} {a_{4}} {a_{5}}} R^{{a_{2}} {a_{3}} {a_{4}} {a_{5}}}-48R^{m {a_{1}}} R^{n {a_{2}} {a_{3}} {a_{4}}} R^{{a_{1}} {a_{2}} {a_{5}} {a_{6}}} R^{{a_{3}} {a_{4}} {a_{5}} {a_{6}}}+192R^{m {a_{1}}} R^{n {a_{2}} {a_{3}} {a_{4}}} R^{{a_{1}} {a_{5}} {a_{3}} {a_{6}}} R^{{a_{2}} {a_{6}} {a_{4}} {a_{5}}}+96R^{m {a_{1}}} R^{n {a_{2}} {a_{1}} {a_{3}}} R^{{a_{2}} {a_{4}} {a_{5}} {a_{6}}} R^{{a_{3}} {a_{4}} {a_{5}} {a_{6}}}+8R^{m n} R^{{a_{1}} {a_{2}} {a_{3}} {a_{4}}} R^{{a_{1}} {a_{2}} {a_{5}} {a_{6}}} R^{{a_{3}} {a_{4}} {a_{5}} {a_{6}}}-32R^{m n} R^{{a_{1}} {a_{2}} {a_{3}} {a_{4}}} R^{{a_{1}} {a_{5}} {a_{3}} {a_{6}}} R^{{a_{2}} {a_{6}} {a_{4}} {a_{5}}}+96R^{n {a_{1}}} R^{m {a_{2}} {a_{1}} {a_{3}}} R^{{a_{2}} {a_{4}} {a_{5}} {a_{6}}} R^{{a_{3}} {a_{4}} {a_{5}} {a_{6}}}-48R^{n {a_{1}}} R^{m {a_{2}} {a_{3}} {a_{4}}} R^{{a_{1}} {a_{2}} {a_{5}} {a_{6}}} R^{{a_{3}} {a_{4}} {a_{5}} {a_{6}}}+192R^{n {a_{1}}} R^{m {a_{2}} {a_{3}} {a_{4}}} R^{{a_{1}} {a_{5}} {a_{3}} {a_{6}}} R^{{a_{2}} {a_{6}} {a_{4}} {a_{5}}}+96R^{m {a_{1}} n {a_{2}}} R^{{a_{1}} {a_{3}} {a_{2}} {a_{4}}} R^{{a_{3}} {a_{5}} {a_{6}} {a_{7}}} R^{{a_{4}} {a_{5}} {a_{6}} {a_{7}}}-48R^{m {a_{1}} n {a_{2}}} R^{{a_{1}} {a_{3}} {a_{4}} {a_{5}}} R^{{a_{2}} {a_{3}} {a_{6}} {a_{7}}} R^{{a_{4}} {a_{5}} {a_{6}} {a_{7}}}+192R^{m {a_{1}} n {a_{2}}} R^{{a_{1}} {a_{3}} {a_{4}} {a_{5}}} R^{{a_{2}} {a_{6}} {a_{4}} {a_{7}}} R^{{a_{3}} {a_{7}} {a_{5}} {a_{6}}}-192R^{m {a_{1}} {a_{2}} {a_{3}}} R^{n {a_{4}} {a_{5}} {a_{6}}} R^{{a_{1}} {a_{5}} {a_{2}} {a_{7}}} R^{{a_{3}} {a_{4}} {a_{6}} {a_{7}}}-96R^{m {a_{1}} {a_{2}} {a_{3}}} R^{n {a_{4}} {a_{5}} {a_{6}}} R^{{a_{1}} {a_{5}} {a_{4}} {a_{7}}} R^{{a_{2}} {a_{3}} {a_{6}} {a_{7}}}-96R^{m {a_{1}} {a_{2}} {a_{3}}} R^{n {a_{4}} {a_{2}} {a_{5}}} R^{{a_{1}} {a_{5}} {a_{6}} {a_{7}}} R^{{a_{3}} {a_{4}} {a_{6}} {a_{7}}}-48R^{m {a_{1}} {a_{2}} {a_{3}}} R^{n {a_{4}} {a_{2}} {a_{3}}} R^{{a_{1}} {a_{5}} {a_{6}} {a_{7}}} R^{{a_{4}} {a_{5}} {a_{6}} {a_{7}}}+192R^{m {a_{1}} {a_{2}} {a_{3}}} R^{n {a_{4}} {a_{2}} {a_{5}}} R^{{a_{1}} {a_{6}} {a_{3}} {a_{7}}} R^{{a_{4}} {a_{6}} {a_{5}} {a_{7}}}-192R^{m {a_{1}} {a_{2}} {a_{3}}} R^{n {a_{4}} {a_{2}} {a_{5}}} R^{{a_{1}} {a_{6}} {a_{4}} {a_{7}}} R^{{a_{3}} {a_{6}} {a_{5}} {a_{7}}}-48R^{m {a_{1}} {a_{2}} {a_{3}}} R^{n {a_{4}} {a_{5}} {a_{6}}} R^{{a_{1}} {a_{7}} {a_{2}} {a_{3}}} R^{{a_{4}} {a_{7}} {a_{5}} {a_{6}}}+96R^{m {a_{1}} {a_{2}} {a_{3}}} R^{n {a_{4}} {a_{5}} {a_{6}}} R^{{a_{1}} {a_{7}} {a_{2}} {a_{4}}} R^{{a_{3}} {a_{7}} {a_{5}} {a_{6}}}+24R^{m {a_{1}} {a_{2}} {a_{3}}} R^{n {a_{1}} {a_{4}} {a_{5}}} R^{{a_{2}} {a_{3}} {a_{6}} {a_{7}}} R^{{a_{4}} {a_{5}} {a_{6}} {a_{7}}}-96R^{m {a_{1}} {a_{2}} {a_{3}}} R^{n {a_{1}} {a_{4}} {a_{5}}} R^{{a_{2}} {a_{6}} {a_{4}} {a_{7}}} R^{{a_{3}} {a_{7}} {a_{5}} {a_{6}}}-96R^{m {a_{1}} {a_{2}} {a_{3}}} R^{n {a_{1}} {a_{2}} {a_{4}}} R^{{a_{3}} {a_{5}} {a_{6}} {a_{7}}} R^{{a_{4}} {a_{5}} {a_{6}} {a_{7}}}+12R^{m {a_{1}} {a_{2}} {a_{3}}} R^{n {a_{1}} {a_{2}} {a_{3}}} R^{{a_{4}} {a_{5}} {a_{6}} {a_{7}}} R^{{a_{4}} {a_{5}} {a_{6}} {a_{7}}} - \frac{1}{2}\delta^{m n} L^{(D,4)}
\end{dmath}


\bibliographystyle{elsarticle-num}
\bibliography{bibliography}

\begin{thebibliography}{10}
\expandafter\ifx\csname url\endcsname\relax
  \def\url#1{\texttt{#1}}\fi
\expandafter\ifx\csname urlprefix\endcsname\relax\def\urlprefix{URL }\fi
\expandafter\ifx\csname href\endcsname\relax
  \def\href#1#2{#2} \def\path#1{#1}\fi

\bibitem{castillo-felisola20_cadab_python_algor_gener_relat_cosmol_ii}
O.~Castillo-Felisola, K.~Peeters, D.~T. Price, M.~Scomparin, Cadabra and python
  algorithms in general relativity and cosmology {II}: {G}ravitational {W}aves
  (2020).

\bibitem{MacCallum:2018csx}
M.~A.~H. MacCallum, {Computer algebra in gravity research}, Living Rev. Rel.
  21~(1) (2018) 6.
\newblock \href {https://doi.org/10.1007/s41114-018-0015-6}
  {\path{doi:10.1007/s41114-018-0015-6}}.

\bibitem{peeters07_cadab}
K.~Peeters, Cadabra: a field-theory motivated symbolic computer algebra system,
  Comput. Phys. Commun. 176~(8) (2007) 550.
\newblock \href {https://doi.org/10.1016/j.cpc.2007.01.003}
  {\path{doi:10.1016/j.cpc.2007.01.003}}.

\bibitem{peeters07_introd_cadab}
K.~Peeters, {Introducing Cadabra: A Symbolic Computer Algebra System for Field
  Theory problems} (2007).
\newblock \href {http://arxiv.org/abs/hep-th/0701238}
  {\path{arXiv:hep-th/0701238}}.

\bibitem{peeters07_symbol_field_theor_with_cadab}
K.~Peeters,
  \href{http://www.fachgruppe-computeralgebra.de/CA-Rundbrief/car41.pdf}{Symbolic
  field theory with cadabra}, Computeralgebra Rundbrief 41 (2007) 16.
\newline\urlprefix\url{http://www.fachgruppe-computeralgebra.de/CA-Rundbrief/car41.pdf}

\bibitem{Peeters:2018dyg}
K.~Peeters, {Cadabra2: computer algebra for field theory revisited}, J. Open
  Source Softw. 3~(32) (2018) 1118.
\newblock \href {https://doi.org/10.21105/joss.01118}
  {\path{doi:10.21105/joss.01118}}.

\bibitem{baran2019analytical}
V.~Baran, D.~Delion, Analytical approach for the quartet condensation model,
  Physical Review C 99~(3) (2019) 031303.

\bibitem{brewin19_using_cadab_tensor_comput_gener_relat}
L.~Brewin, \href{http://arxiv.org/abs/1912.08839v1}{Using cadabra for tensor
  computations in general relativity} (2019).
\newblock \href {http://arxiv.org/abs/1912.08839} {\path{arXiv:1912.08839}}.
\newline\urlprefix\url{http://arxiv.org/abs/1912.08839v1}

\bibitem{einstein15_zur_allgem_relat}
A.~Einstein, Zur allgemeinen relativit{\"a}tstheorie, Sitzungsber. Preuss.
  Akad. Wiss. 1 (1915) 778.

\bibitem{einstein16_grund_allgem_relat}
A.~Einstein, Die grundlage der allgemeinen relativit\"atstheorie, Ann. Phys.
  49~(4) (1916) 284--339.

\bibitem{cadabra_man}
K.~Peeters, \href{https://cadabra.science/man.html}{Cadabra manual pages}.
\newline\urlprefix\url{https://cadabra.science/man.html}

\bibitem{lanczos38_remar_proper_rieman_chris_tensor_four}
C.~Lanczos, {A Remarkable Property of the Riemann--Christoffel Tensor in Four
  dimensions}, Annals Math. (1938) 842.

\bibitem{lovelock69_uniquen_einst_field_equat_four_dimen_space}
D.~Lovelock, The uniqueness of the einstein field equations in a
  four-dimensional space, Archive for Rational Mechanics and Analysis 33~(1)
  (1969) 54.

\bibitem{lovelock71_einst_tensor_its}
D.~Lovelock, {The Einstein Tensor and Its generalizations}, J. Math. Phys. 12
  (1971) 498--501.
\newblock \href {https://doi.org/10.1063/1.1665613}
  {\path{doi:10.1063/1.1665613}}.

\bibitem{padmanabhan.13_lancz_lovel_model_gravit}
T.~Padmanabhan., D.~Kothawala., Lanczos-lovelock models of gravity 531~(3)
  (2013) 115--171.
\newblock \href {https://doi.org/10.1016/j.physrep.2013.05.007}
  {\path{doi:10.1016/j.physrep.2013.05.007}}.

\bibitem{mueller-hoissen85_spont_compac_with_quadr_cubic_curvat_terms}
F.~Mueller-Hoissen, Spontaneous compactification with quadratic and cubic
  curvature terms 163 (1985) 106--110.
\newblock \href {https://doi.org/10.1016/0370-2693(85)90202-3}
  {\path{doi:10.1016/0370-2693(85)90202-3}}.

\bibitem{mueller-hoissen90_from_chern_simon_to_gauss_bonnet}
F.~Mueller-Hoissen, From {Chern-Simons} to {Gauss-Bonnet}, Nucl. Phys. B 346
  (1990) 235--252.
\newblock \href {https://doi.org/10.1016/0550-3213(90)90246-A}
  {\path{doi:10.1016/0550-3213(90)90246-A}}.

\bibitem{verwimp89_higher_dimen_gravit}
T.~Verwimp, On higher dimensional gravity: the lagrangian, its dimensional
  reduction and a cosmological model, Class. Quantum Grav. 6 (1989) 1655--1663.
\newblock \href {https://doi.org/10.1088/0264-9381/6/11/018}
  {\path{doi:10.1088/0264-9381/6/11/018}}.

\bibitem{jebsen21}
J.~T. Jebsen, Ark. Mat. Ast. Fys. 15~(18) (1921).

\bibitem{birkhoff23_relat_moder_physic}
G.~D. Birkhoff, Relativity and Modern Physics, Harvard University Press, 1923.

\bibitem{alexandrow23}
W.~Alexandrow, Ann. der Phys. 72 (1923) 141.

\bibitem{eisland25}
J.~Eisland, Trans. Amer. Math. Soc. 23 (1925) 213.

\bibitem{weinberg08_cosmol}
S.~Weinberg, Cosmology, Oxford, 2008.

\bibitem{brewin19_hybrid-latex}
L.~Brewin, hybrid-latex, \url{https://github.com/leo-brewin/hybrid-latex}
  (2019).

\end{thebibliography}





\end{document}